%% file: FDESL.tex
\documentclass[11pt]{article}
\usepackage{amsthm,amsmath,natbib}
\RequirePackage[colorlinks,citecolor=blue,urlcolor=blue]{hyperref}
\usepackage{amsfonts}
\usepackage{amssymb}
\usepackage{amsmath}
\usepackage{natbib}
\usepackage{fontenc}
\usepackage{undertilde}
\usepackage{amsfonts}
\usepackage{epsfig}
\usepackage{latexsym,graphics,graphicx,amsbsy,amsmath,amsthm}
\usepackage[table]{xcolor}
\usepackage{setspace}
\usepackage{soul}
\usepackage{rotating}
\usepackage{algpseudocode}
\usepackage{algorithm}
\usepackage{caption}
\usepackage{slashbox}
\usepackage{authblk}

\newtheorem{remark}{Remark}[section]

\RequirePackage[OT1]{fontenc}
\RequirePackage[colorlinks]{hyperref}

\newcommand{\lam}{\lambda}
\newcommand{\cd}{\overset{\cal D}{\rightarrow}}

\begin{document}

\title{\textbf{On the use of the cumulant generating function for inference on time series}}

\author{A. Moor\thanks{Corresponding author: alban.moor@unige.ch, Bd Pont-d’Arve 40, CH-1211 Geneva 4, Switzerland.}, D. La Vecchia, E. Ronchetti \thanks{Research Center for Statistics and Geneva School of Economics and Management, University of Geneva.}}

\date{\today}
\maketitle
\begin{abstract}
We introduce innovative inference procedures for analyzing time series data. Our methodology enables density approximation and composite hypothesis testing based on Whittle’s estimator, a widely applied M-estimator in the frequency domain. Its core feature involves the (general Legendre transform of the) cumulant generating function of the Whittle likelihood score, as obtained using an approximated distribution of the periodogram ordinates. We present a testing algorithm that significantly expands the applicability of the state-of-the-art saddlepoint test, while maintaining the numerical accuracy of the saddlepoint approximation. Additionally, we demonstrate connections between our findings and three other prevalent frequency domain approaches: the bootstrap, empirical likelihood, and exponential tilting. Numerical examples using both simulated and real data illustrate the advantages and accuracy of our methodology.
\end{abstract}

\vspace{0.5cm}

\vspace{0.5cm}

\section{Introduction}

Short length time series are common in  many scientific areas. In general, small time series datasets arise
 when sampling is costly or when there are structural changes over the considered period and one has to slice the sample in subperiods. For instance, series containing only few hundreds of data are routinely encountered in computational economics (see, e.g., \citet{IN05}),  macroeconomics (\citet{NP82, S92, LVR19}); 
business analytics (see, e.g., 
\citet{MH03} and \citet{AH11}); biology (see e.g., \cite{LD10}); ecology (\cite{Be95});  climatology (\cite{Mu10}). Other examples are available in the literature about bootstrap for time series. 

The analysis of time series datasets containing a small number of records represents a statistical challenge: in small samples the  commonly applied 
asymptotic inferential methods typically perform poorly. Indeed, the first order asymptotic approximation tends to be inaccurate in the tails of the distribution (even for moderate sample sizes), 
which is usually the area of interest for inference (e.g. for the computation of $p$-values). Additionally, the approximation 
typically works well and is theoretically justified if the  sample size diverges to infinity, but often its accuracy  
deteriorates quickly in moderate to small samples. Thus, the finite and small sample performance of inferential methods is
a crucial operational aspect in the daily practice of time series data analysis. The development of techniques designed to perform well
even when the number of observations is less than few hundreds can be beneficial for many scientific areas: it will yield more accurate inference, which in turn will enable reliable decision making.

\subsection{Motivating example} \label{Sec: MotivEx}

To illustrate numerically the central problems of the extant asymptotics, let us consider the flexible and widely applied class of autoregressive fractionally integrated moving average (ARFIMA) processes; see among others \citet{R03} and \citet{Beranetal13} for a book-length introduction. We work on an ARFIMA$(p,d,q)$ process, having dynamics 
\begin{equation}
\label{Eq: FARIMAG}
\phi(L)(1-L)^{d}X_t =\theta(L) \epsilon_t,
\end{equation} 
where $L$ is the back-shift operator $LX_t=X_{t-1}$, $\phi(L)$ is the autoregressive (AR) polynomial of order $p$, $\theta(L)$ is the moving average (MA) polynomial  of order $q$, and, $\forall t \in \mathbb{Z}$, the $\{\epsilon_t\}$ are i.i.d.\ with zero mean and known variance $\sigma^2=1$. 
When $d=0$, we recover the standard ARMA$(p,q)$ model, whilst for $d=1$, we have the ARIMA$(p,1,q)$. For values of $d\in(0,1/2)$ we have a long memory process, whose autocovariance function has an hyperbolic decay when the number of lags diverges, or, equivalently, whose spectral density has a pole at the origin. 

Let us focus on an ARFIMA of order $p=2$, $d=0$ and $q=0$ process, with AR coefficients  $\phi_1= 0.1$, $\phi_2= 0.2$, and Gaussian errors with unit variance. The model parameter is $(\phi_1,\phi_2,d)$. We consider the sample sizes $n=250,2500,5000$, and estimate 
the parameter via the frequency domain approach: we make use of the widely applied Whittle's estimator (\cite{W53}), as implemented in the routine \texttt{WhittleEst} available in the R package \texttt{longmemo}. To test 
$\mathcal{H}_0: d = 0 \quad \text{vs} \quad \mathcal{H}_1: d > 0,$
(namely, we test for the presence of long memory) we resort on the Wald test statistic---see (\ref{wald}). For each considered value of $n$, we repeat the exercise 2500 times (Monte Carlo runs) and obtain the distribution of the Wald statistic in each setting. The routinely applied first order asymptotic theory implies that for large sample sizes the Wald statistic has a $\chi_1^2$-distribution,  under  $\mathcal{H}_0$. 

\begin{figure}[h!]
\begin{center}
\begin{tabular}{ccc}

 $n=250$ & $n=2500$ & $n=5000$\\  
\includegraphics[width=0.3\textwidth, height=0.275\textheight, angle=0]{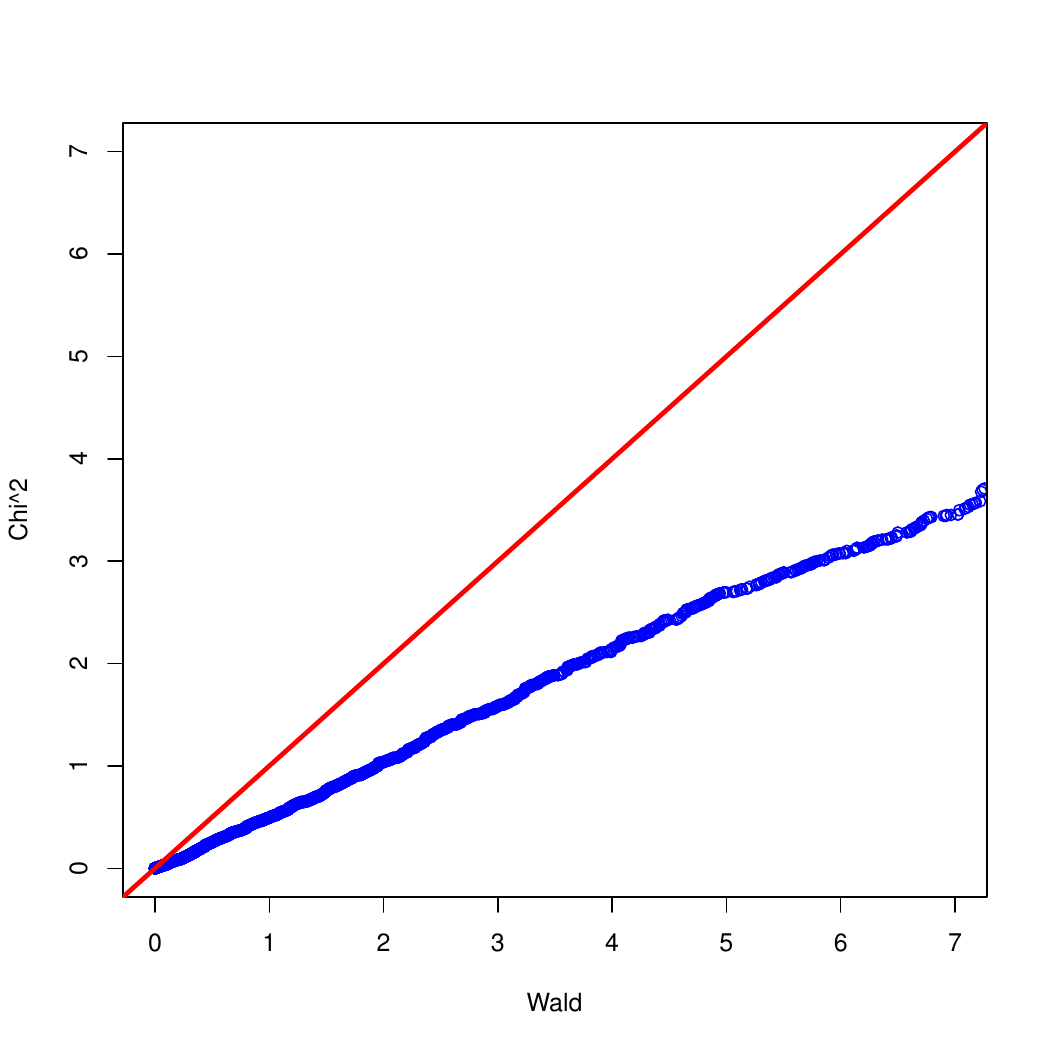}\ &
\includegraphics[width=0.3\textwidth, height=0.275\textheight, angle=0]{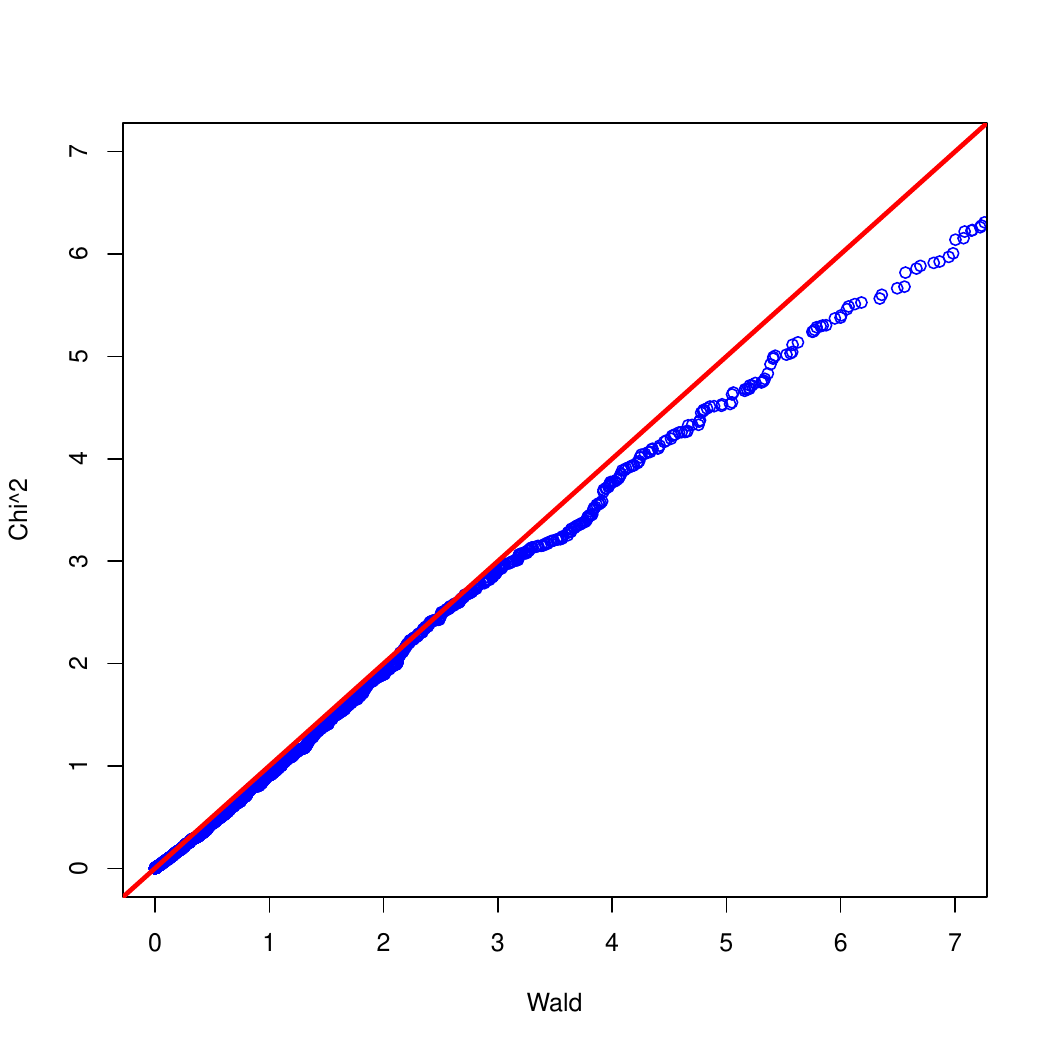} & 
\includegraphics[width=0.3\textwidth, height=0.275\textheight, angle=0]{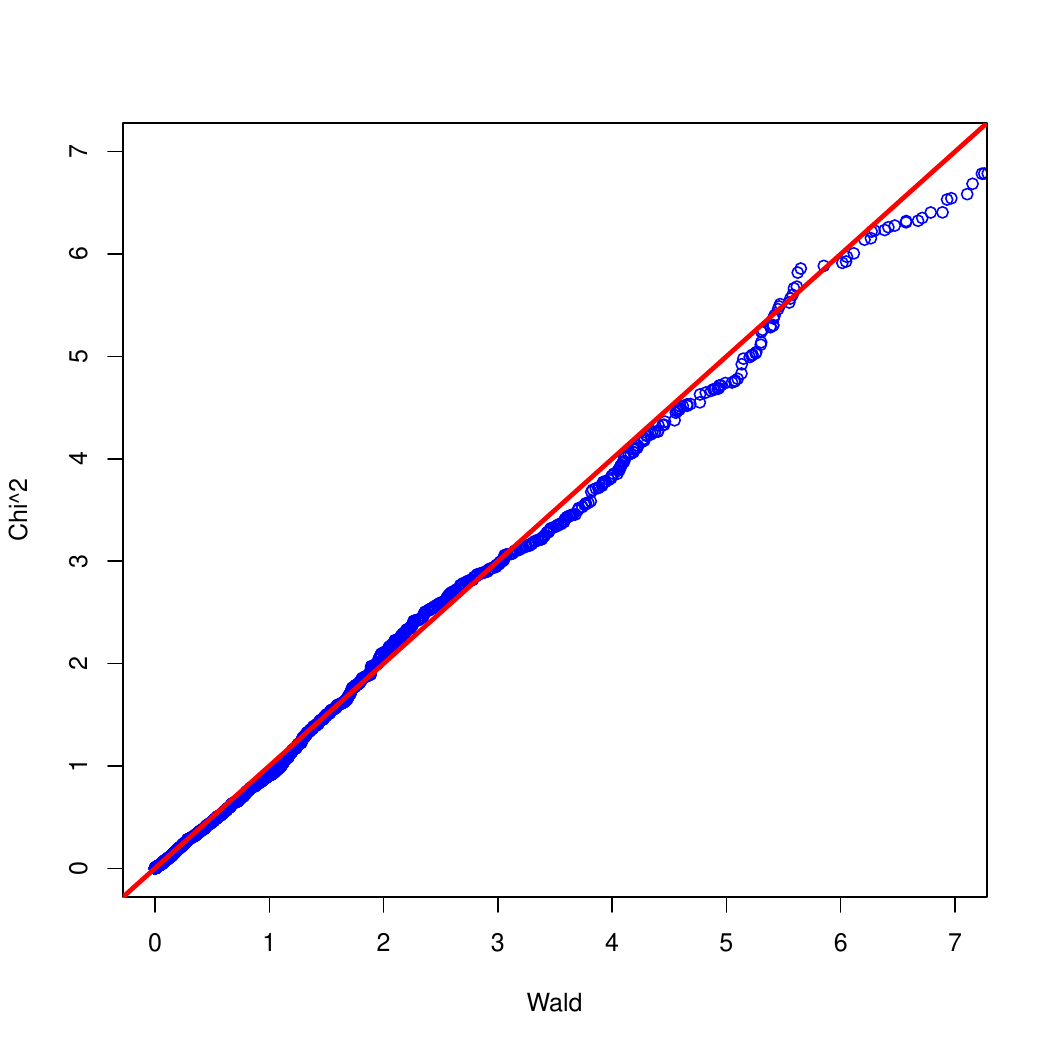}\\
\end{tabular} 
\end{center}

\caption{ARFIMA$(2,d,0)$ model. Wald-type test statistics, for $\mathcal{H}_0:\ d=0$ versus $\mathcal{H}_1: \ d > 0$. For different sample sizes $n$, the QQ-plots compare the quantiles of the asymptotic  $\chi^2_1$-distribution to the actual distribution (as obtained using 2500 Monte Carlo runs) of the Wald statistic.}
    \label{Fig: QQWald}
\end{figure}

In Figure \ref{Fig: QQWald}, we display the QQ-plot of the Wald statistic for different values of $n$. The plots illustrate the inaccuracy of the first order asymptotic approximation for $n=250$, with accuracy improvements being appreciable from  $n=2500$. For instance, considering the moderately large sample size $n=2500$, the actual 95-percentile of the Wald statistic is about 4.22, whilst the first order asymptotic approximated chi-square 95-percentile is about 3.84: the asymptotic approximation entails a relative error (see e.g.\ \cite{LVRI23} p.38 for a definition) of about $9.0\%$.  Still focusing on the 95-percentile, the relative error decreases to about $6.0\%$ when $n=5000$: doubling the sample size reduces the relative error by only one third. Finally, we notice that for any considered sample size, the relative error grows as we move deeper into the right tail: for instance, for the 99-percentile the relative error is about 11.2\% when $n=2500$.

\subsection{Related work}

The motivating example casts doubt on the accuracy (and ultimately the usefulness) of tests based on the available asymptotic theory already 
for samples sizes of the order of few hundreds. To cope with this inference issue, several \textit{higher order (or small sample) asymptotic techniques} have been developed. These techniques include 
the bootstrap, the Edgeworth and the saddlepoint; we refer to \cite{Y09} and references therein for a review in the independent and
identically distributed (i.i.d.) setting.

 In the Gaussian ARFIMA process considered in the motivating example, \cite{LA05} derive the Edgeworth expansion for Whittle's estimator density: this tool can be applied for testing
$\mathcal{H}_0$ e.g. by means of the $p$-value obtained by numerical integration of the approximated density.
However, using the first terms of this expansion provides in general a 
good approximation in the center of the density, but it can be inaccurate in the tails, where it 
can even become negative. It is well-known in the statistical literature that 
saddlepoint techniques overcome these problems: they are accurate in the tails (sample space asymptotics, see \cite{WBB93})
and perform well also in small samples  
(sample size asymptotics). We refer to  the seminal paper of \cite{D54} and, for book-length presentations, to
\cite{FR90}, \cite{Jensen95}, \cite{K06}, \cite{butler2007}, and  \cite{brazzaleetal2007}, among others.

The theory of saddlepoint techniques is well developed in the case of i.(i).d.\ observations. In contrast, 
only a few results have been obtained in the time series setting. 
The available techniques are essentially developed in the time domain and they consider 
the saddlepoint density approximations for the first 
order partial correlation coefficient in autoregressive processes (AR) of order one
with 
Gaussian errors; see \cite{daniels1956}, \cite{P78}, \cite{D80},
\cite{W92}, \citet{M11} and \cite{FR13}. Only recently, \citet{LVR19} introduced saddlepoint density approximations and 
tests in the presence of nuisance parameters
for ratio statistics and Whittle's estimator. These techniques hinge on the semi-parametric 
assumption 
that the standardized periodogram ordinates are i.i.d. exponentially distributed, for any sample size $n$ and not only for large $n$. We label them as 
\textit{exponential-based saddlepoint techniques}. The resulting approximations
perform well for both short- and long-memory time series.

\subsection{Our contributions}

In this paper, we build on the theoretical results of \citet{LVR19} and we introduce novel saddlepoint techniques (density, tail-area approximations, and testing in the presence of nuisance parameters)  for
Whittle's estimator.

 Our techniques complement the toolkit already available in the literature on saddlepoint approximations. The key aspect of our developments is that we drop the assumption of exponential distribution of the periodogram ordinates, which  is central in \citet{LVR19}. Specifically, we explain how to define saddlepoint techniques relying on the {empirical distribution} 
(rather than the theoretical asymptotic exponential distribution) of the periodogram ordinates. Our construction builds on the intuition of \citet{RW94}, who define
the empirical saddlepoint density approximation in the setting of independent data. We label the resulting techniques as \textit{frequency domain empirical saddlepoint} (FDES), and develop an algorithm allowing to use them efficiently to test composite hypotheses. FDES techniques are computationally easier to implement than the ones in \citet{LVR19}, since they approximate integrals by their sample counterparts. In parallel, our testing methodology is handily workable in comparison to the state-of-the-art saddlepoint test of \citet{RRY03}, while preserving a better accuracy than the first order asymptotics.

Overall, our methodological contribution is 
fourfold. (i) We derive novel saddlepoint techniques using the empirical distribution of the periodogram ordinates, treated as independent random variables (Section \ref{FDESdef}).  (ii) We connect our approach to other extant frequency domain techniques that have already exploited the empirical distribution of the periodogram ordinates: the frequency domain
bootstrap (FDB) of \citet{DJ96},  the empirical likelihood (FDEL) of \citet{M97}, and the exponential tilting (FDET) of \cite{kakizawa13}. Building on these papers and on \citet{RM93}, we identify some connections between our FDES, the FDB, the FDEL and the FDET (in Section \ref{connect}). (iii) The central aspect of our methodology is the approximation of $p$-values for frequency domain test statistics. This approximation is based on the numerical integration (via importance sampling) of the saddlepoint density approximation to the exact sampling distribution of Whittle's estimator, which improves on the routinely applied $p$-values approximation, as obtained using the first order asymptotic Gaussian density. We provide a pseudo-code containing the key steps needed for the implementation our procedure (in Section \ref{TestMV}). (iv) Finally, we conduct numerical experiments (on simulated and real data) to illustrate the advantages and the accuracy of our procedure for popular time series models (Section \ref{Section: MC}).

\section{Setting, definitions and first order asymptotics} \label{Sec: setting}

Let $\{X_t\}$ be a linear and second order stationary process, with spectral density function indexed 
by a parameter $\theta$
\begin{equation}
\label{Eq: gen_spec}
f(\lambda; \theta) = \vert \lambda \vert^{-2d} L(\lambda;\utilde{\vartheta}),	
\end{equation}
where  $\lambda \in \Pi = (-\pi, \pi ]$, $d \in [0,0.5)$, $\utilde{\vartheta} \in \mathbb{R}^{p-1}$ and $\theta=(d \
\utilde\vartheta)\in\Theta\subset\mathbb{R}^p$, with  $p \geq 1$ and  $\Theta$ is a compact set. The function $L(\cdot;\utilde{\vartheta})$ in (\ref{Eq: gen_spec})
is even, bounded on every compact subinterval of $(0, \pi ]$ and slowly varying at zero, that is,
$$\lim_{\lambda\rightarrow 0} L(\lambda a;\utilde{\vartheta})/L(\lambda;\utilde{\vartheta}) = 1$$
for all $a \in (0,\infty)$.

Following \citet{L03}, we classify the process $\{X_t\}$ as short-range dependent (SRD) or long-range dependent (LRD) based 
on the  parameter $d$ 
and the behavior of $L(\cdot;\utilde{\vartheta})$
near the origin. 
Indeed, when $d = 0$ and the function 
$L(\cdot;\utilde{\vartheta})$ is bounded 
with $L(0;\utilde{\vartheta})\neq 0$, the process $\{X_t\}$ features SRD. Otherwise, the process $\{X_t\}$ features LRD;
we refer to \citet{R03} for a book-length presentation. 

We consider the problem of conducting inference on the model parameter which indexes the spectral density. We consider that 
Assumptions A-E in \cite{LVR19}
hold. Therefore, we assume that there exists a zero mean, variance $\sigma^2_\xi$, and zero third cumulant $k_3$ process, say $\{\xi_t\}$, of i.i.d.\ random variables,   
such that $\{X_t\}$ has 
the one-sided representation:
$X_t =\sum_{r=0}^{\infty} a_r \xi_{t-r},$ 
for $t\in \mathbb{Z}$ and $\{a_r\}$ is a sequence of constants 
satisfying either $\sum_{r=0}^\infty \vert a_r \vert < \infty$ or $a_r \sim L_a(r) r^{d-1}$, 
with $L_a(\cdot)$ being slowly varying at infinity. According to our parametrization, $\sigma^2_\xi$ is therefore an element of $\utilde{\vartheta}$.

Given $n$ observations $X_1,X_2,...,X_n$ and the model defined in (\ref{Eq: gen_spec}), 
we perform a transformation aimed to weaken the dependence in the data.
Specifically, let $h : [0, 1] \rightarrow \mathbb{R}$ be a function of bounded variation. Then, the tapered discrete 
Fourier transform (DFT)  is 
\begin{equation}
\label{DFT_tap}
d_{n} (\lam) = \sum_{t=1}^{n} h \left( \frac{t}{n}\right) X_t \exp{(-it\lam)} , \text{ \ \ \ $\lam \in \Pi$},
\end{equation}
where $i^2 = {-1}$  and the data-taper $h(\cdot)$ satisfies Assumptions 5 and 6 in \cite{DJ96}. A data taper is typically used for handling missing data or to reduce the leakage (e.g., caused by a time series behaviour close to non-stationarity); see, e.g., \citet{T68},  \citet{D88} or \cite{Br81} and \citet{B04}. The non-tapered version of the DFT, is obtained by setting $h\equiv 
1$. The periodogram ordinate at $\lam$ is  
\begin{equation}\label{Eq: Per}
 I(\lambda)=  \frac{\vert d_{n} (\lam) \vert^2 }{{2\pi \sum_{t=1}^{n} h^2 \left({t}/{n}\right)}}.  
\end{equation}

Now, 
let us first assume that  $\sigma^2_\xi=1$. We specify a parametric model for the spectral density $f(\lam;\theta)$, so 
the (negative) Whittle's log-likelihood for $\theta \in \Theta \subset [0,0.5) \times \mathbb{R}^{p-1}$ 
with $p\geq 1$,  
is 
\begin{equation} \label{Eq: WhittleLik}
\mathcal{L}_W(\theta) = \frac{1}{2\pi}
\Big\{ \int_{\Pi} \ln f(\lambda;\theta) d\lambda +
 \int_{\Pi}  \frac{I(\lambda)}{f(\lambda;\theta)} d\lambda \Big\}. 
\end{equation}
The likelihood in \eqref{Eq: WhittleLik} can be applied for estimation and for hypothesis testing.

In terms of estimation, the optimization of (\ref{Eq: WhittleLik}) 
defines Whittle's estimator as the
root of $\nabla_\theta \mathcal{L}_W(\theta)=0$, that is 
\begin{equation} \label{Eq: WhittleCont}
\int_{\Pi}  \left(\frac{I(\lambda)}{f(\lam;\theta)} - 1 \right) \frac{\nabla_{\theta}f(\lam;\theta)}{f(\lam;\theta)} d\lambda =0,
\end{equation}
we refer to \cite{LVR19} for additional details.

To implement (\ref{Eq: WhittleCont}), we consider the discrete frequency form of the 
Whittle's likelihood which is simply obtained  by a common Riemann-type discretization. 
Specifically, the DFT is evaluated at 
Fourier frequencies
$\lam_{j,n} = 2 \pi \left({j}/{n}\right),$ for $j=1,2,...m$ and $m=\lfloor (n-1)/2 \rfloor$, 
and the periodogram ordinate at $\lam_{j,n}$ is defined by $ I(\lambda_{j,n})$.

We write  $\lambda_j=\lambda_{j,n}$, $I_j=I(\lambda_{j})$ and, for $j=1,2,...m$, the Riemann approximation of (\ref{Eq: 
WhittleCont}) is 
\begin{equation}
\sum_{j=1}^m \psi_j(I_j;\hat{\theta}_n) = 0 \ , 
\label{Eq: Mestimator}
\end{equation}
where  $\psi_j: \mathbb{R}^{+} \times \mathbb{R}^{p} \rightarrow \mathbb{R}^{p}$ has the form
\begin{equation}
\psi_j\left(I_j;\theta\right)=\left(\frac{I_{j}}{f(\lambda_j;\theta)} -1 \right) z_j(\theta),
\label{Eq: psiM}
\end{equation}  
with  $z_j(\theta) = \nabla_{\theta} \ln f(\lambda_j;\theta) $ computed at $\lambda_j$ and $z_j(\theta) \in \mathbb{R}^p$. 

Letting $\theta^{0}$ be the true parameter value, the first order asymptotic distribution of Whittle's
estimator can be derived treating the standardized periodogram ordinates as independent and identically 
distributed r.v.s, having exponential density with rate one:
\begin{equation}
I_j/f(\lambda_j;\theta^0) \sim   \chi_2^2/2.
\label{Eq. exp1}
\end{equation}

\noindent Under the same treatment and correct specification of the spectral density, each $\psi_j$ in (\ref{Eq: psiM}) has expectation zero. Thus, it defines a set of estimating equations which yield a frequency domain M-estimator.

We refer to \citet{Beranetal13} for the mathematical details of the  first order asymptotic theory of Whittle's estimator. Here, we simply sketch the ideas hinging on standard results on M-estimators (see e.g. \cite{vdW98}) and introduce the key quantities for our development. 

A first order Taylor 
expansion 
yields:

\begin{equation}
\sqrt{n}(\hat{\theta}_n-{\theta}^0) \approx \left[-\frac{1}{n}\sum_{j=1}^{m} \nabla_{\theta} \psi_j\left(I_j;{\theta}^0\right)\right]^{-{T}} 
\left( \frac{1}{\sqrt{n}}  \sum_{j=1}^{m} \psi_j\left(I_j;{\theta}^0\right) \right). \label{Eq. WAsy}
\end{equation}
The estimating function $\psi_j$ is linear in $I_j/f(\lambda_j;\theta^0)$, thus (\ref{Eq. exp1}), the central limit theorem and the weak law of large numbers imply that
$\sqrt{n}(\hat{\theta}_n-{\theta}^0) \cd \mathcal{N}\left(0, \Sigma_{W} (\theta^0)\right),$
where the asymptotic variance is obtained using $\Sigma_W(\theta) = 4 \pi  \left(\int_{\Pi} \left( \nabla_\theta \ln f (\lambda;\theta)\right) \left( \nabla_\theta \ln f (\lambda;\theta) \right)^T d\lam\right)^{-1}$.

If $\sigma^2_\xi$ is unknown, we need to estimate it and the above construction applies with some modifications. Specifically, we  notice that Whittle's likelihood 
can be rewritten by choosing a special scale-parameterization: setting  $\theta^*=(\sigma^2_{\xi} \ \vartheta)$ and $\theta=(1 \ \vartheta)$,  

the likelihood in (\ref{Eq: WhittleLik}) becomes:
\begin{equation*}
\mathcal{L}_W(\vartheta) = \sum_{j =1}^m  \frac{I_j}{f(\lambda_{j};\theta)}.
\end{equation*}

Minimizing $\mathcal{L}_W(\vartheta)$
is equivalent to obtain the frequency domain M-estimator $\hat\vartheta_n$ solution to
\begin{equation}
\sum_{j =1}^m \frac{\nabla_{\vartheta}f(\lambda_j;\theta)}{f(\lambda_j;{\theta})} 
\left(\frac{I_j}{f(\lambda_j;{\theta^*})} \right)=0, 
\label{Eq: Mestimator2}
\end{equation}
which is the Riemann discretization (as obtained using the Fourier frequencies) of 
\begin{equation} \label{Eq: WhittleCont2}
\int_{\Pi} \frac{\nabla_{\theta}f(\lam;\theta)}{f(\lam;\theta)} \left(\frac{I(\lambda)}{f(\lam;\theta)} \right) d\lambda = 0.
\end{equation}

Once $\hat\vartheta_n$ is available, we estimate the remaining parameter $\sigma^2_{\xi}$ by $\hat{\sigma}^2_{\xi,n} = \mathcal{L}_W(\hat\vartheta_n)/m$.
The resulting estimator  $ \hat\theta^*_n=(\hat{\sigma}^2_{\xi,n} \ \hat\vartheta_n)$ is asymptotically normal.

As far as the hypothesis testing problem is concerned,
\begin{equation}
\mathcal{H}_0: \theta = \theta^0 \quad \text{vs} \quad \mathcal{H}_1: \theta \neq \theta^0, 
\label{Eq. TestProblem}
\end{equation}
the first order Gaussian asymptotic provides the theoretical underpinning for the definition of the Wald test statistic (asymptotically equivalent to
the Whittle likelihood ratio statistic)

which is, under $\mathcal{H}_0$ in \eqref{Eq. TestProblem}, asymptotically distributed as a $\chi^{2}_{p}$.

The motivating example in \S \ref{Sec: MotivEx} illustrates the poor finite sample performance of the first order asymptotics for $\theta$, when testing hypothesis on $d$. \citet{LVR19} show that the use of saddlepoint techniques  improve on the first order asymptotics. Their construction is based on a Legendre-type transform of the cumulant generating function (henceforth c.g.f.), a key tool that we introduce in the next section.
 
\section{Saddlepoint techniques}

{To introduce our new tools (see Section \ref{Sec: FDESadd}), first we need to
 recall some results in \cite{LVR19} (see Section \ref{Sect: Exp_Sadd}). The key point of our argument is that
 one can obtain saddlepoint techniques in the frequency domain using the approximated
 c.g.f.\ of the score function of Whittle's estimator. The techniques discussed in Section \ref{Sect: Exp_Sadd} and Section \ref{Sec: FDESadd} differ in the
 way in which the approximated c.g.f. is obtained, but they make use of the same tool: the 
 general Legendre transform of the approximated c.g.f.}

\subsection{Exponential-based techniques} \label{Sect: Exp_Sadd}

Saddlepoint density approximations and saddlepoint based testing procedures in the frequency domain were introduced in
\cite{LVR19}. These techniques are derived treating the periodogram ordinates as independent random variables and they fully exploit the asymptotic theory of DFT based on the exponential distribution
of the periodogram ordinates given in \eqref{Eq. exp1}.
The resulting approximation is higher-order accurate for SRD processes, first order accurate for LRD, and it is analytically tractable;
see Propositions 3.1 and 3.2 in  \cite{LVR19}. 

The key ingredient in the density approximation 
is
the approximated c.g.f of the score function $\psi_j(I_j; \theta)$ of Whittle's estimator, that La Vecchia and Ronchetti obtain as
\begin{eqnarray} 
\tilde K_{\psi_j} (\upsilon; \theta) &=& \ln \left(\tilde{E} \left[ \exp\{\upsilon^T \psi_j(I_j; \theta)\}  \right]\right) \label{Eq: kumulant} \label{Eq: Kpsij} \\
&=&\ln\left(-\frac{1}{ f(\lambda_j;\theta^0)\tilde{s}_j} \right)-\upsilon^T z_j(\theta), \nonumber 
\end{eqnarray} 
where $\upsilon \in \mathbb{R}^{p},$ $\tilde E[\cdot]$ is the expected value computed under (\ref{Eq. exp1}), and

\begin{equation*}
\tilde{s}_j(\theta)= \frac{1}{f(\lambda_j;\theta)} \upsilon^T z_j(\theta) - \frac{1}{f(\lambda_j;\theta^0)}; 
\end{equation*}
{for more details, see (3.25) in the supplementary material of \cite{LVR19} and Appendix \ref{App: FEXP} for an example about
Fractional Exponential Processes (FEXP)}. 

Testing can also be performed using saddlepoint based techniques by exploiting (\ref{Eq. exp1}) and (\ref{Eq: kumulant}).
To see it, let us consider the problem of simple hypothesis testing (\ref{Eq. TestProblem}).  \citet{LVR19} show how to adapt to 
time series setting the saddlepoint test statistic defined in \citet{RRY03} for i.i.d. data. The proposed test statistic is:
\begin{equation} \label{Eq. SADtest1}
\tilde S(\hat \theta_n) = 2 \mathcal{\tilde K}^{\dagger}(\hat\theta_n),
\end{equation}
where, for $\mathcal{\tilde {K}}(\upsilon;\theta)=\sum_{j=1}^m \tilde K_{\psi_j} (\upsilon; \theta)$,
\begin{eqnarray}
\mathcal{\tilde K}^{\dagger}(\theta) &=& \sup_{\upsilon} \{-\tilde{\mathcal K}(\upsilon;\theta)\}   
               = -\tilde{\mathcal K}(\upsilon_0(\theta);\theta),  \label{Eq. Lpsi} 
\end{eqnarray}
where $\upsilon_0 = \upsilon_0(\theta)$ solves the saddlepoint equation
\begin{equation}
\partial_\upsilon \mathcal{\tilde K} (\upsilon;\theta) = 0.
\end{equation}
\citet{LVRI23} call  $\mathcal{\tilde K}^{\dagger}$ the general Legendre transform of $\mathcal{\tilde K}$ {and study some of its  statistical and mathemathical properties}.

The distribution of 
$\tilde S(\hat \theta_n)$ under the null, can be approximated by a $\chi_{p}^{2}$. 
This approximation is obtained by integrating the saddlepoint density approximation. The resulting approximation typically remains accurate down to small sample sizes, and the test is asymptotically (first order) equivalent to the Wald test of \S \ref{Sec: MotivEx}, but has better small sample properties; see \cite{LVR19} for a discussion.

In the daily practice of time series data analysis, simple hypotheses are the exception rather than the rule. The case of composite hypotheses is a practically more relevant problem and it is related to the problem of performing hypothesis testing in the presence of nuisance parameters. The motivating example provides a typical situation where the statistician is interested in testing on $d$, while the other ARFIMA parameters are 
nuisance parameters. Also in this challenging case, $\mathcal{\tilde K}^{\dagger}$ is the central tool needed to perform such a test.

To develop further, let us define a partition $(\theta_{(1)} \ \theta_{(2)})$ of the original parameter $\theta,$ and test hypothesis only on the first components $\mathcal{H}_0 : \theta_{(1)}=\theta_{(1)}^0$ w.l.o.g., with $\theta_{(2)}$ treated as nuisance parameters. Whittle's estimator is $\hat{\theta}_n=(\hat \theta_{(1),n} \ \hat \theta_{(2),n})$ and \cite{LVR19} show that the saddlepoint test statistic defined by \citet{RRY03} can be adapted to the time series setting. Specifically, the test statistic is obtained by concentrating out the nuisance parameters in (\ref{Eq. Lpsi}):
\begin{eqnarray}
\tilde S(\hat \theta_{(1),n}) 
  &=& 2\inf_{\theta_{(2)}}  \mathcal{\tilde K}^{\dagger}( \hat \theta_{(1),n} \ \theta_{(2)} )     \label{Eq. SADtest}   \\
  &=& 2\inf_{\theta_{(2)}} \left[ \sup_{\upsilon}\left\{ -\sum_{j=1}^{m}
                      \tilde{K}_{\psi_j}(\upsilon_;(\hat \theta_{(1),n} \ \theta_{(2)} ))\right\}\right].  \nonumber
 \end{eqnarray}
The test statistic is an univariate quantity which takes care of the nuisance parameters by a minimax optimization---via the infimum in \eqref{Eq. SADtest}. Under the null, its exact distribution can be approximated by a $\chi_{p_1}^{2}$,
where $p_1=\text{dim}(\theta_{(1)})$; see \citet{LVR19}.

From the computational standpoint,
the test is attractive since it does not require any integration for the marginalization of the nuisance parameters;
see \citet{JR94} and \citet{RRY03}.
However, in our experience, the need to perform a maximization ($ \sup_{\upsilon}$) nested in a minimization ($\inf_{\theta_{(2)}}$) can
make the evaluation of $S(\hat \theta_{(1),n})$ computationally challenging. To cope with this issue, we propose a numerical procedure that takes care of the nuisance parameters in a different way, avoiding the concentration $\inf_{\theta_{(2)}}$ in the case of composite hypotheses testing (see Remark \ref{Remark1}). The central aspect of this novel methodology is the approximation of $p$-values based on the numerical integration (via importance sampling) of the saddlepoint density approximation to the exact sampling distribution of Whittle's estimator; see Section \ref{TestMV}.

\subsection{Empirical saddlepoint techniques} \label{Sec: FDESadd} 
As mentioned in \citet{TDCM12}, the frequency domain approach to time series analysis is particularly appealing because the use of the periodogram ordinates reduces a dependent data problem (the analysis of time series data in the time domain) to an independent data one (the analysis of asymptotically independent periodogram ordinates). \cite{M97} exploits this property to define FDEL confidence regions for Whittle's estimator in the SRD setting. Her construction relies on the empirical distribution of
the periodogram ordinates, treated as independent random variables.

Therefore Monti's approach provides the basic intuition to derive saddlepoint techniques for time series using an alternative method to the one summarized in Section \ref{Sect: Exp_Sadd}.
Namely, rather than resorting on the asymptotic exponential distribution of the periodogram ordinates, we move along the same lines as \citet{DJ96} and \citet{M97} and we define  saddlepoint
techniques making use of {the empirical distribution of the periodogram ordinates}. We label the resulting tools as \textit{Frequency Domain Empirical Saddlepoint (FDES)} techniques. These techniques are based on the theory and method developed for FDEL in \citet{M97} combined  with the construction derived in \citet{RW94}. Also for the FDES, we are going to illustrate the pivotal role played by the general Legendre transform of the empirical c.g.f. \\

\subsubsection{Density approximation}\label{FDESdef}  

To begin with, let us consider the FDES density approximation for the multivariate parameter estimate $\hat\theta_n$, as obtained by Whittle's  method. 
Let $P$ be the actual distribution of $I_j/f(\lam_j;\theta^0)$, not necessarily an exponential distribution. Then,  
the $I_j/f(\lam_j;\theta^0)$
are independent and identically distributed and
there exists a saddlepoint $\upsilon_0=\upsilon(\theta^0)$ defined by 
\begin{equation}
\int_{\mathbb{R}^{+}} \psi_j(I_j;\theta^0) \exp\{\upsilon_0^T\psi_j(I_j;\theta^0)\} dP=0, \label{Eq. Lemma}
\end{equation} 
which has continuous derivative  in a neighborhood of $\theta^0$ and satisfies $\upsilon(\theta^0+m^{-1/2} u) = O(m^{-1/2})$ uniformly for 
$u$ in a compact set; 
see \cite{RW94}. Making use of this result,  we define the empirical saddlepoint density approximation for Whittle's estimator $\hat{\theta}_n$ at a generic point $\theta$ belonging to the so-called normal region $(\hat{\theta}_n-m^{-1/2}u,\hat{\theta}_n+m^{-1/2}u)$, for $u$ in a compact set. Then, for $p=\text{dim}(\Theta)$, the empirical saddlepoint density approximation is
\begin{equation}
\hat{g}_{\hat{\theta}_n}(\theta)=\left(\frac{m} {2\pi} \right)^{p/2}\left\vert \det \hat{M}(\theta)\right\vert \left\vert \det\hat{\Sigma}(\theta)\right\vert
^{-1/2} \exp\{m \hat{K}(\theta)\},
\label{Eq: empiricalSAD}
\end{equation}
where
\begin{equation}
\hat{K}
(\theta) 
= \ln \left[   \frac{1} {m} \sum_{j=1}^{m} \exp\{{\hat\upsilon^T} \psi_{j}(I_j;\theta) \}\right],
\label{Eq: Kemprictxt}
\end{equation}
\begin{equation*}
\hat{M}
(\theta) 
=  \exp\{-\hat{K}(\theta)\}  \frac{1}{m} \sum_{j=1}^{m} \nabla_{s}\psi_{j}(I_j;s)\vert_{s=\theta}  \exp\{\hat\upsilon^T \psi_{j}(I_j;\theta) \}, 
\label{Eq: Mempric}
\end{equation*}
\begin{equation*}
\hat{\Sigma}
(\theta) 
=  \exp\{-\hat{K}(\theta)\} \frac{1}{m} \sum_{j=1}^{m}  \psi_{j}(I_j;\theta) \psi_{j}(I_j;\theta)^T \exp\{\hat\upsilon^T \psi_{j}(I_j;\theta) \}, 
\label{Eq: Sempric}
\end{equation*}
and the empirical saddlepoint  $\hat\upsilon$ (for ease of notation, occasionally we write $\hat \upsilon$ instead of
 $\hat \upsilon(\theta)$) satisfies:
\begin{equation}
\sum_{j=1}^{m} \psi_{j}(I_j;\theta)\exp \{\hat\upsilon^T\psi_{j}(I_j;\theta)\}  =0.
\label{Eq: aempiric1}
\end{equation}
If we define
\begin{equation}
\hat{K}(\upsilon;\theta) 
= \ln \left[   \frac{1} {m} \sum_{j=1}^{m} \exp\{{\upsilon^T} \psi_{j}(I_j;\theta) \}\right],
\label{Eq: K_upsilon_theta}
\end{equation}
solving \eqref{Eq: aempiric1} is equivalent to finding $\sup_{\upsilon} \{-\hat{K}(\upsilon;\theta)\}$, namely $\partial_\upsilon \hat{K}(\upsilon;\theta) =0$. Thus, similarly to \eqref{Eq. Lpsi}, we define
\begin{eqnarray}
\hat{K}^\dagger(\theta) &=& \sup_{\upsilon} \{-\hat{K}(\upsilon; \theta)\}  \label{Eq. empdagger}  \\
                                  &=& -\hat{K}(\hat{\upsilon}(\theta);\theta) = -\hat{K}(\theta)  \nonumber
\end{eqnarray}
and we can express the empirical saddlepoint density approximation using the general Legendre transform of the empirical c.g.f.\ as:
$$
\hat{g}_{\hat{\theta}_n}(\theta)=\left(\frac{m} {2\pi} \right)^{p/2}\left\vert \det \hat{M}(\theta)\right\vert \left\vert \det\hat{\Sigma}(\theta)\right\vert
^{-1/2} \exp\{- m \hat{K}^{\dagger}(\theta)\}.
$$

Letting $\theta^0$ denote the true parameter value, we have that $f_{\hat\theta_n}(\theta^0+m^{-1/2}u)$, namely the actual density of Whittle's estimator evaluated at $\theta^0+m^{-1/2}u$,
can be approximated by 
\begin{eqnarray*}
\begin{small}
\hat{g}_{\hat\theta_n}(\hat\theta_n + m^{-1/2}u) =  
 \left(\frac{m} {2\pi} \right)^{p/2}\left\vert \det \hat{M}(\hat\theta_n + m^{-1/2}u)\right\vert \left\vert \det\hat{\Sigma}(\hat\theta_n + m^{-1/2}u)\right\vert
^{-1/2} e^{m\hat{K}(\hat\theta_n + m^{-1/2}u)}.
\end{small}
\end{eqnarray*}
Moving along the lines  of \cite{RW94} and treating the periodogram ordinates as independent (see \cite{M97}),
one can prove that,  for any $u$ in a compact set,
$$
f_{\hat\theta_n}(\theta^0+m^{-1/2}u)=\hat{g}_{\hat\theta_n}(\hat\theta_n + m^{-1/2}u)\{1+m^{-1/2}a_m(u)+O_P(m^{-1})\},
$$ 
as $m \rightarrow \infty$, where $a_m(u)=O_P(1)$ is defined in \citet{RW94}. 
  
Just to fix the ideas, let us consider the case of a univariate parameter---the case of numerical integration for a multivariate parameter is discussed in \S \ref{TestMV}. To implement $\hat{g}_{\hat{\theta}_n}$, the first step is to obtain $\hat\theta_n$ and to determine the support of the sampling distribution. Then,  we  define a grid $\left\{v_a: a=1, \ldots, A\right\}$ about $\hat\theta_n$ and we compute the empirical saddlepoint $\hat{\upsilon}(\hat\theta_n+v_a)$, at each grid point. This can be accomplished by noting that, at $v=0$, we have that $\hat{\upsilon}(\hat\theta_n)=0$ and, for each other grid point $v_a$ we need to solve \eqref{Eq: aempiric1}, finding the root of 
$$
 \sum_{j=1}^m \psi_j(\hat\theta_n+v_a) \exp\{\hat{\upsilon}(\hat\theta_n+v_a)^{T} \psi_j(\hat\theta_n+v_a)\}=0.
$$
These equations can be solved e.g. by Newton-Raphson algorithm (or a secant method), with initial value $\hat{\upsilon}(\hat\theta_n+v_{a-1})$---namely, we make use of the value of the saddlepoint at the previous grid point. In our experience, for smooth spectral densities (which imply the smoothness of $\psi_j$) and for a sufficiently fine grid (e.g. $A=100$), this procedure is accurate and computationally fast. Once the sequence of saddlepoints is available, we compute $\hat{K}(\hat\theta_n+v_a), \hat{M}(\hat\theta_n+v_a), 
\hat{\Sigma}(\hat\theta_n+v_a)$ and then compute $\hat{g}_{\hat{\theta}_n}(\hat\theta_n + v_a)$ at each grid point. As it is customary in the saddlepoint approximation literature, we suggest to normalize the resulting approximation, the normalizing constant being 
$$C_a= \sum_{a=1}^A\left(v_a-v_{a-1}\right) \hat{g}_{\hat\theta_n}(\hat\theta_n + v_a).$$

The FDES density approximation has a connection to  the FDB. Indeed, the bootstrap procedure described in \citet{DJ96}, on p. 1937, can be obtained in two 
ways. Either one relies on standard exponentially distributed random variables for the periodogram ordinates, or one makes use of the empirical distribution of the periodogram ordinates. We remark that our FDES techniques are, by construction,  connected to this latter approach. Therefore, in the normal region, formula (\ref{Eq: empiricalSAD}) yields results which are similar to those obtained using the bootstrap density. However, the FDES approximation does not require resampling; see 
\citet{DH88} and \citet{RW94} for a related discussion. This is a computational advantage over the FDB: our empirical saddlepoint techniques do not require to solve Whittle's estimating equation for each bootstrap sample. This feature not only makes the implementation of the empirical saddlepoint density faster than the one obtained by FDB, but it also avoids some computational issues which affects the FDB. Indeed, in our experience, the computation of Whittle's estimator can be problematic: for sample sizes $n \lesssim 150$ the routine can fail to find an estimator. The FDES avoids these numerical issues, as it only requires the computation of the original estimator $\hat{\theta}_n$.

Finally, in the i.i.d.\ setting, we notice that \citet{HolcblatSowell2022} use (\ref{Eq: empiricalSAD}) as an Empirical Likelihood (EL) and maximize
$\ln[\hat{g}_{\hat{\theta}_n}(\theta)] $ with respect to $\theta$ to obtain a new estimator called
empirical saddlepoint estimator. Furthermore, 
\citet{FasioloWoodHartigBravington2018} use the empirical saddlepoint approximation to approximate the density of
summary statistics in synthetic likelihoods.

\subsection{Connection to empirical likelihood and exponential tilting}\label{connect}

Note that our saddlepoint is based on the c.g.f.\ $\hat{K}(\upsilon;\theta)$ in \eqref{Eq: K_upsilon_theta} as an approximation 
to the true c.g.f. Its general Legendre transform is the key tool needed to compute $\hat{g}_{\hat\theta_n}$ and it unveils two important
connections with the generalized empirical likelihood in the frequency domain.
Specifically, we illustrate how $\hat K^\dagger(\theta)$ 
allows us to connect the FDES with the EL test statistic 
proposed in \citet{M97} and with the results about FDET in \cite{kakizawa13}. 

To begin with, we recall that \citet{RM93} show that the empirical saddlepoint has a
connection to the EL, in the i.i.d.\ setting. Following those arguments, we first bridge our FDES density approximation and the FDEL.
To this end, following \cite{M97}, we notice that the FDEL solves 
the system of (tilted) estimating equations 
\begin{equation}
\sum_{j=1}^{m} \psi_{j}(I_j,\theta) [1+ \hat\xi^T\psi_j(I_j;\theta)]^{-1} =0,
\label{Eq: EL}
\end{equation}
where  we use the shorthand notation $\hat\xi=\hat\xi(\theta)$. Then, Monti defines Owen's statistics as
\begin{equation}
\hat{W}(\theta)= 2 \sum_{j=1}^{m} \ln\{1+ \hat\xi^T \psi_j(I_j;\theta) \} 
\label{WaldFD}
\end{equation}
and shows that $\hat{W}(\theta) =  \tilde{W}(\theta) + O_P(n^{-1/2}),$ where 
$\tilde{W}(\theta)$ is the Wald-type statistic defined as 
\begin{align}\label{wald}
\tilde{W}(\theta) = m (\hat\theta_n-\theta)^{T}  \hat{V}^{-1} (\hat\theta_n-\theta),
\end{align}
with $\hat V = \hat A^{-1}(\hat\theta_n)\hat B(\hat\theta_n) \hat A^{-T}(\hat\theta_n)$, 
$$
\hat A(\hat\theta_n)= \frac{1}{m}   \sum_{j=1}^{m} \nabla_{\theta}\psi_{j}(I_j;\theta) \big\vert_{\theta=\hat\theta_n} \; \text{and} \; \hat{B}(\hat\theta_n) 
=  \frac{1}{m}  \sum_{j=1}^{m}  \psi_{j}(I_j;\hat\theta_n) \psi_{j}(I_j;\hat\theta_n)^T. 
$$

Now, to get some insight into the link between $\hat\xi$ and $\hat\upsilon$, let us compare (\ref{Eq: aempiric1}) to (\ref{Eq: EL}). From (\ref{Eq. Lemma}), it follows that $\upsilon(\theta)=O(n^{-1/2})$, for $\theta$ in a root-$n$ neighbourhood of $\theta^0$. Then, a Taylor expansion in $\upsilon$ yields
$\exp\{\upsilon^{T} \psi_j(I_j;\theta)\}=1 + \upsilon^T \psi_j(I_j;\theta) + O_P(n^{-1})$ (see Appendix \ref{AppC}). Thus, considering \eqref{Eq: aempiric1}, the saddlepoint satisfies the equation 
$$\sum_{j=1}^{m} \psi_{j}(I_j;\theta) [1+ \hat\upsilon^T\psi_j(I_j;\theta)] =O_P(n^{-1}).$$ 
Then, another Taylor expansion of the equation defining the FDEL  yields 
$$\sum_{j=1}^{m}\psi_j(I_j;\theta) [1+ \hat\xi^T\psi_j(I_j;\theta)]^{-1}=\sum_{j=1}^{m} \psi_{j}(I_j;\theta) [1- \hat\xi^T\psi_j(I_j;\theta)] =O_P(n^{-1}),
$$ 
since $\hat\xi=O_P(n^{-1/2})$. Thus, we conclude that the empirical saddlepoint and the empirical likelihood solve at the order 
$O_P(n^{-1})$ the same equation. 

Equipped with this result, in Appendix \ref{AppC} we illustrate
how to connect Monti's expression of Owen's statistics 
$\hat{W}(\theta)$
to $\hat K(\theta)$. In particular, we 
combine the results in \citet{RM93} and in \citet{M97} to obtain 
\begin{equation}
- {2}n \hat K(\theta) = 2n \hat K^\dagger(\theta) = 2\hat W(\theta) - \frac{2m^{-1/2}}{3} \sum_{j=1}^{m} \left\{ u^T \hat M^T \hat\Sigma^{-1} \psi_j(I_j;\hat\theta_n)   \right
\}^3 + R_n
\label{Eq. MYFDEL}  
\end{equation}
where 
$\hat{\Sigma} = \hat{\Sigma} (\hat\theta_n)$ and  $\hat{M} = \hat{M}(\hat\theta_n)$. Treating the periodogram ordinates as independent, the term $R_n$ in (\ref{Eq. MYFDEL}) is of order $O_P(n^{-1})$ as in \cite{RM93}; see \cite{M97} p.400
for a related discussion. Thus,  (\ref{Eq. MYFDEL}) represents a frequency domain analogous of the result in \citet{RM93} and it is a novel finding in the literature on higher-order asymptotics for time series. 

Equation (\ref{Eq. MYFDEL}) has a threefold importance: (i) it yields a nonparametric approximation of the density of Whittle's estimator based on the FDEL; (ii) it illustrates that difference between the statistics based on $\hat K^\dagger(\theta)$ and on $\hat W(\theta)$ depends on the skewness of the Whittle's score; incidentally, it shows that they both correct the Wald statistic taking into account the skewness. (iii) it allows to connect our FDES to the FDEL.

Among these tree points, we illustrate that the last one is of help in hypothesis testing. To this end, we recall that \cite{kakizawa13}  studies a class of frequency domain generalized EL  methods and provide the large sample theory of several test statistics for parametric restrictions. Kakizawa's theory covers the FDEL and a version of the FDET as special cases and it is based on the i.i.d.\ treatment of the standardized periodogram ordinates. This hints at the possibility of also connecting our FDES (in particular the $\hat K^\dagger(\theta) $) to both Wald and FDET type of test statistics.

\noindent A Taylor expansion yields: 
\begin{eqnarray}
\hat{K}^\dagger(\theta^0) = \hat{K}^\dagger(\hat{\theta}_n)  - \hat{K}^{\dagger \prime}(\hat{\theta}_n)^{T}(\hat{\theta}_n - \theta^0) + (\hat{\theta}_n - \theta^0)^{T}\hat{K}^{\dagger \prime \prime}(\hat{\theta}_n)(\hat{\theta}_n - \theta^0)/2 + O_P(n^{-3/2}).
\label{Alb2}
\end{eqnarray}
\noindent By definition of $\hat \theta_n$ in \eqref{Eq: Mestimator} and of the empirical saddlepoint in \eqref{Eq: aempiric1},
 we have $\hat \upsilon(\hat \theta) = 0,$ which implies $\hat{K}^\dagger(\hat{\theta}_n) = \hat{K}(0;\hat \theta_n) = 0$ and $\hat{K}^{\dagger \prime}(\hat{\theta}_n) = \sum_{j=1}^{m} \psi_{j}(I_j;\hat \theta_n) = 0.$ Thus, 
\begin{align}\label{QuadFHat}
2n\hat{K}^\dagger(\theta^0) = n(\hat{\theta}_n - \theta^0)^{T}\hat{K}^{\dagger \prime \prime}(\hat{\theta}_n)(\hat{\theta}_n - \theta^0) + O_P(n^{-1/2}).
\end{align}
\noindent The left-hand side of \eqref{QuadFHat} is similar to the Exponential Tilting (ET) statistic; see e.g.\ \citet{kitamura_ET_1997} (Equation (10)) and \citet{ISJ98} (Equation (9)) in the i.i.d.\
setting.

The first order asymptotic theory implies that $2n\hat{K}^\dagger(\theta^0),$  has a $\chi^2_{p}$ distribution. Alternatively, to approximate the distribution of $\hat{K}^\dagger(\theta^0),$ and thus the level of the test, we propose to use our empirical saddlepoint approximation $\hat{g}_{\hat\theta_n}(\theta)$ (as in (\ref{Eq: empiricalSAD})) to the density of $\hat{\theta}_n$ under $\mathcal{H}_0.$ To this end, we define the Wald-type statistic
$$
\tilde{W}_n (\theta)= n (\hat\theta_n-\theta)^{T} \hat{V}^{-1} (\hat\theta_n-\theta)
$$
and we obtain 
$$
2n\hat{K}^\dagger(\theta^0) = \tilde{W}_n (\theta^0) + O_P(n^{-1/2}).$$

Using \eqref{Eq. WAsy}, the distribution of $\tilde{W}_n (\theta)$ is approximated by the first order $\chi_p^2$. With the aim of improving numerically the accuracy approximation of the $p$-value,  we propose to make use of $\hat{g}_{\hat\theta_n}(\theta)$ 
 as
\begin{align}\label{pVal}
\mathbb{P}[\tilde{W}_n(\theta^0) > \tilde{w}(\theta^0) \mid \mathcal{H}_0] \approx \int_{ \mathcal{B} } \hat{g}_{\hat\theta_n}(\theta) d\theta,
\end{align}
\noindent where  
$\tilde{w}(\theta)$ is the observed value of the test statistic and 
$$
 \mathcal{B} = \left\{ \theta \in \mathbb{R}^d \mid  \tilde{w}(\theta) > \tilde{w}(\theta^0) \right\}.
$$

\subsection{Hypothesis testing with FDES} \label{TestMV}

The implementation of the approach in  \eqref{pVal} requires the computation of the integral defining the $p$-value. We suggest to use Monte Carlo methods, which usually outperform deterministic numerical methods in multidimensional setups (see e.g.\ \cite{CRMC} for a book-length introduction). 

More specifically, we recommend to use an importance sampling scheme based on an instrumental Gaussian distribution, which makes use of the information available in Whittle's estimator. The idea goes as follows: if we were able to sample directly from our FDES approximation $\hat{g}_{\hat\theta_n}(\theta),$ we could rely on the Strong Law of Large Numbers to approximate $\int_{ \mathcal{B} } \hat{g}_{\hat\theta_n}(\theta) d\theta$ 
with any desired degree of accuracy. As this is generally impossible, our strategy is to generate sample points from the Gaussian $ Z \sim \mathcal{N}(\hat \theta_n, \hat{V}),$ and correct the deviation of the sample using weights based on the FDES $\hat{g}_{\hat\theta_n}$, via:
$$
\int_{ \mathcal{B} } \hat{g}_{\hat\theta_n}(\theta) d\theta = \int_{\Theta} \frac{\mathbb{I}_\mathcal{B}(\theta)\hat{g}_{\hat\theta_n}(\theta) \phi_{\hat \theta_n, \hat{V}}(\theta)}{\phi_{\hat \theta_n, \hat{V}}(\theta)}d\theta = {E}_{\phi_{\hat \theta_n, \hat{V}}}\left[\frac{\mathbb{I}_\mathcal{B}(Z )\hat{g}_{\hat\theta_n}(Z)}{\phi_{\hat \theta_n, \hat{V}}(Z)}\right], 
$$
\noindent where $\mathbb{I}_\mathcal{B}(\cdot)$ is the indicator function,  $\phi_{\mu,\Sigma}$ is the Gaussian density with mean $\mu$ and covariance matrix $\Sigma$, and the
expectation is estimated by averaging over the Monte Carlo (MC) samples.

The advantage of this approach is that it exploits the knowledge of the (first order asymptotic) Gaussian distribution, which is a natural reference distribution for the MC procedure, with the advantage of being readily available in every statistical software. We itemize the key steps of the FDES hypothesis testing procedures in the pseudo-codes available in Algorithm \ref{algo1} below. We give the more fundamental importance sampling procedure in Algorithm \ref{algoIS}. This numerical approach based on the FDES represents an alternative procedure to the $\chi^2$-approximation obtained in \cite{RRY03} and mentioned in \S \ref{Sect: Exp_Sadd}.

\begin{algorithm}[h!]
\caption{Testing the simple hypothesis $\mathcal{H}_0 : \theta=\theta^0.$ The function $\operatorname{mvrnorm}(\cdots)$ generates data from a multivariate Gaussian with parameters $\mu$ and $\Sigma,$ and we define the function $\operatorname{ImpSample}(\cdots)$ in Algorithm \ref{algoIS}, where $R$ is the MC sample size.}\label{algo1}
  \begin{algorithmic}[1]
	\Require {$X_1,\ldots, X_n,$ $R,$ $\theta^0$ }
						\State $\left\{I_n(\lambda_i)\right\}_{i=1}^{n} \leftarrow \lvert \operatorname{FFT}(X_1,\ldots, X_n) \rvert^2$ \Comment{Use $O( n \ln n)$ complex Fast Fourier Transform.}
						\State $m \leftarrow \lfloor (n-1)/2 \rfloor$
            \For {$j = 1,\ldots,m$}
						\State $\psi_j\left(I_j;\theta\right) \leftarrow \left(\frac{I_{j}}{f(\lambda_j;\theta)} -1 \right) z_j(\theta)$ \Comment{Define the Whittle's score as in \eqref{Eq: psiM}}
						\EndFor
						\State $\hat{\theta}_n \leftarrow \operatorname{Solve}_{\theta} \left( \sum_{j=1}^m \psi_j\left(I_j;\theta\right) = 0 \right)$ 
						\State $\hat{g}_{\hat\theta_n}(\theta) \leftarrow \left(\frac{m} {2\pi} \right)^{\text{dim}(\Theta)/2}\left\vert \det \hat{M}(\theta)\right\vert \left\vert \det\hat{\Sigma}(\theta)\right\vert^{-1/2} \exp\{m \hat{K}(\theta)\}$ \Comment{Define the saddlepoint density as in \eqref{Eq: empiricalSAD}.}
						\State  $\mathcal{B} \leftarrow \left\{ \theta \in \Theta \mid \tilde{w}(\theta) > \tilde{w}(\theta^0) \right\}$ 
						\Procedure{ISintegral}{$\mathcal{S}$}
						\State $\operatorname{ImpSample}(\operatorname{function} = \hat{g}_{\hat\theta_n}, \operatorname{set} = \mathcal{S}, \operatorname{generator} = \operatorname{mvrnorm}(\mu=\hat{\theta}_n,\Sigma = \hat{V}),\operatorname{size} = R)$
						\EndProcedure
						\State $p \leftarrow \operatorname{ISintegral}(\mathcal{B})/\operatorname{ISintegral}(\Theta)$
            \State \textbf{return} $p$ \Comment{The $p$-value as in \eqref{pVal}.}
    \end{algorithmic}
\end{algorithm}

\begin{algorithm}[h!]
\caption{Function $\operatorname{ImpSample}$ : computes an integral $\int_{ \mathcal{S} } f(x) dx$ by importance sampling. The function $\operatorname{mvrnorm}(\cdots)$ generates data from a multivariate Gaussian with parameters $\mu$ and $\Sigma,$ and the function $\operatorname{mvdnorm}(\cdots)$ gives the corresponding Gaussian density.}\label{algoIS}
  \begin{algorithmic}[1]
	\Require {$\operatorname{function} = f, \operatorname{set} = \mathcal{S}, \operatorname{generator} = \operatorname{mvrnorm}(\mu,\Sigma), \operatorname{size}=R$}
	\For{$r = 1,\ldots,\operatorname{R}$}
	\State $z_r \leftarrow \operatorname{mvrnorm}(\mu,\Sigma)$
	\State $x_r \leftarrow \mathbb{I}_\mathcal{S}(z_r) f(z_r) / \operatorname{mvdnorm}(z_r,\mu,\Sigma)$
	\EndFor
	\State \textbf{return} $\sum_{r=1}^{R} x_r /\operatorname{R}$
  \end{algorithmic}
\end{algorithm}

We present here the FDES testing hypothesis for the Wald-type statistic only, but the same procedure remains valid with other test statistics, as long as they are (at least approximately) functions of $\hat{\theta}_n$. Examples of such test statistics, as the FDEL and FDET, are available in Section \ref{connect}.
Indeed, the proposed numerical procedure is a very versatile tool, which can be used to obtain accurate $p$-values for any test statistic endowed with a saddlepoint density approximation (either the exponential-based one or its empirical version).  Moreover, it can be applied to obtain confidence regions by inverting the test.  In addition to our own numerical experiments (Section \ref{Section: MC}), this method has already proven to be fast and accurate in the i.i.d.\ setting (see e.g.\ \cite{BB2008}). While other MC methods are also usable (e.g. rejection sampling or MCMC methods), importance sampling strikes a good balance between time complexity, simplicity of programming and accuracy, for the moderate dimensions typically encountered
in the analysis of time series data (i.e.\ $p \lesssim 10^2$). 
The evaluation of a single MC sample of, say, $R$ points actually allows to approximate the entire cumulative distribution function (c.d.f.) of $\tilde{W}_n(\theta^0),$ as it readily embeds the (inherently unknown) scaling constant of the saddlepoint $\hat{g}_{\hat\theta_n}(\theta)$. To do so, one can simply change $\tilde{w}(\theta^0)$ to any $x \in \mathbb{R}^{+}$ in \eqref{pVal} and Line 8 of Algorithm \ref{algo1}. Moreover, the i.i.d.\ nature of the MC sample points also allows for faster parallel computing.

\begin{remark} \label{Remark1} 
There are cases where only certain components of $\theta$ have to be tested. Namely, taking the same partition $(\theta_{(1)} \ \theta_{(2)})$ as in Section \ref{Sect: Exp_Sadd}, we test $\mathcal{H}_0 : \theta_{(1)}=\theta_{(1)}^0$ w.l.o.g. To take into account the effect of estimating the nuisance parameter, we marginalize the FDES density with respect to $\theta_{(2)}$. This integration does not affect the computational complexity of our algorithm and does not slow down its implementation. Indeed, testing in the presence of nuisance parameters can be handily achieved with a modification of Line 8 in Algorithm \ref{algo1}, redefining the integration set as
$$
\mathcal{B} \leftarrow \left\{ \theta \in \Theta \mid \tilde{w}\left( \theta_{(1)},\hat{\theta}_{(2),n} \right) > \tilde{w}\left(\theta_{(1)}^0,\hat{\theta}_{(2),n}\right) \right\},
$$ 
instead of 
$$
\mathcal{B} \leftarrow \left\{ \theta \in \Theta \mid \tilde{w}(\theta) > \tilde{w}(\theta^0) \right\}.
$$
As aforementioned, this is different from (\ref{Eq. SADtest}): our numerical integration via importance sampling avoids to take the infimum w.r.t.\ to the nuisance parameters.
\end{remark}

\section{Monte Carlo experiments: ARFIMA process} \label{Section: MC}

Let us consider a ARFIMA$(p,d,q)$ process as   
in (\ref{Eq: FARIMAG}). Let us recall that the spectral density of the  ARFIMA$(p,d,q)$ process at $\lam \in \Pi$ is
\begin{equation}
\label{Eq. FARIMAsp}
f(\lam;d,\eta,\sigma^2)= \frac{\sigma^2}{2 \pi} \frac{|\theta\{\exp (-i \lam)\}|^2}{|\phi\{\exp (-i \lam)\}|^2} \vert 1-\exp(i\lam)  \vert^{-2d}.
\end{equation}

If the variance is known, then we use $\psi_j$ as in (\ref{Eq: psiM}) and our saddlepoint techniques (density and testing can be applied). If the variance $\sigma_\xi^2$ is unknown, we consider it as a nuisance parameter. If the underlying innovation density is Gaussian,  our saddlepoint techniques remain valid. Otherwise, our FDES allows us to easily deal with $\sigma_\xi^2$  by a profiling approach: we concentrate it out from the Whittle likelihood (optimizing w.r.t.\ $\sigma_\xi^2$) as in \citet{M97}. We label  (with a slight abuse of notation) the resulting profiled Whittle scores, for $j=1,...,m$:
\begin{equation}
\psi_j(\eta ) =\frac{I\left(\lam_j\right)}{f_{1}\left(\lam_j;\eta \right)}\left[\frac{\partial \ln \left\{f_{1}\left(\lam_j;\eta \right)\right\}}{\partial \eta }-n^{-1} \sum_{j=1}^n \frac{\partial \ln \left\{f_{1}\left(\lam_j;\eta \right)\right\}}{\partial \eta}\right].
\label{Eq. profiled}
\end{equation}

\subsection{Behaviour of the periodogram ordinates}
\label{Sec: skeptical}

The frequency domain saddlepoint techniques in \S \ref{Sect: Exp_Sadd} are derived under the treatment of  the standardized periodogram ordinates as i.i.d. exponential
for a finite number of Fourier frequencies. One possible source of concern is how realistic 
is this treatment  for small sample sizes. To shed light on this point, we carry out a Shapiro-Wilks test using the following steps on a ARFIMA$(0,d,0)$, with different
innovation densities:

\begin{itemize}
\item[Step 1:]  For the process in (\ref{Eq: FARIMAG}), we consider different sample sizes $n=30, 90, 120$ and values of the long memory parameter:
$d=(0,0.1,0.2,0.25,0.45)$; 
\item[Step 2:] for each specified innovation density, we fix a value of $d$, we simulate a trajectory of length $n$ and for each simulated trajectory, we compute the standardized peridogram ordinates $I_j/f(\lam_j;d)$ and obtain 
$\tilde\epsilon_j=P(I_j/f(\lam_j;d)\leq\epsilon)$, where $P$ is the c.d.f.\ of the exponential distribution with rate one;
\item[Step 3:] we compute $\epsilon_j=\Phi^{-1}(\tilde\epsilon_j)$, where $\Phi^{-1}$ is the inverse c.d.f.\ of the standard Normal;
\item[Step 4:] we apply the Shapiro-Wilk test to $\epsilon_j$ and we see if the test rejects the assumption of normality for $\epsilon_j$, with level $0. 
05$ -- we select such a test since it is well known that it has good 
power even down to small sample sizes;
\item [Step 5:] we repeat the exercise 5000 times, for each value of $d$ and $n$.
\end{itemize}

If the standardized periodogram ordinates are exponential with mean 1, the number of rejections of the normality 
assumption should be close to the 0.05 (nominal level of the Shapiro-Wilk test). In table \ref{Tab: scettico} we show the results. For almost any considered value of $d$, any  sample size and innovation density, the assumption of i.i.d. exponential assumption on periodogram ordinates seems to be 
fairly reasonable. Nevertheless, we remark that when $d=0.45$ the assumption becomes less appropriate for all considered distributions: this is certainly due to the fact that the ARFIMA process is near to non stationarity. Similar considerations apply for  sample sizes larger than $n=120$.

\begin{table}[h!]
\begin{center}
\begin{tabular}{cccccccccccccccc}
\hline \hline
& & &  \multicolumn{4}{c}{Gaussian } &  & & &  \multicolumn{3}{c}{Uniform} \\
& $d$ & & $n = 30$ & & $n= 90$ & & $n= 120$  &&   $n = 30$ & & $n= 90$ & & $n= 120$ \\ \hline
& 0  && 0.038  &&   0.036   &&  0.043 &&  0.033  &&  0.044   &&  0.047\\
& 0.1 && 0.039  &&  0.040    &&  0.047 &&  0.044 &&    0.041 &&  0.047 \\
& 0.2 && 0.034  &&   0.041   &&  0.054 && 0.030  &&   0.049  &&   0.065 \\
& 0.25  && 0.033 &&    0.046 &&    0.062 && 0.030 &&    0.049 &&    0.065 \\
& 0.45  && 0.030 &&    0.109 &&    0.187 &&  0.032 &&    0.110 &&    0.181 \\ \\
\hline \hline \\
& & &  \multicolumn{4}{c}{Student's $t_6$} &  & & &  \multicolumn{3}{c}{Re-centered $\chi^2_5$} \\
& $d$ & & $n = 30$ & & $n= 90$ & & $n= 120$  &&   $n = 30$ & & $n= 90$ & & $n= 120$ \\ \hline
& 0     && 0.037  &&   0.038   &&  0.047 &&  0.037  &&  0.042   &&  0.049\\
& 0.1  && 0.039  &&  0.041    &&  0.049 &&  0.043 &&    0.044 &&  0.047 \\
& 0.2  && 0.036  &&   0.035   &&  0.054 &&  0.031  &&   0.046  &&   0.058 \\
& 0.25  && 0.036 &&    0.040 &&    0.058 && 0.028 &&    0.049 &&    0.062 \\
& 0.45  && 0.030 &&    0.098 &&    0.161 &&  0.027 &&    0.101 &&    0.165 \\

\hline \hline
\end{tabular}
\end{center}
\caption{ARFIMA process: rejection frequency of the Shapiro-Wilks test  (at level $0.05$) for different values of $d$, different sample sizes and innovation densities. Monte Carlo simulation with size $5000$. }
\label{Tab: scettico}
\end{table}

As
FDES techniques of \S \ref{Sec: FDESadd} are concerned, a possible doubt is that  the results in Table \ref{Tab: scettico} may not be appropriate, since to perform the Shapiro-Wilk test, we make use of the c.d.f.\ of the exponential distribution; see Step 2. This criticism has been already addressed in the FEDL of \citet{M97}, who supports the treatment of the periodogram ordinates as independent (without the use of the exponential c.d.f.) by performing  a $\chi^2$-test for independence for any couple $(I_j, I_q)$, $j,q =1,2,... ,n$, under different distributions of the innovation term, in an ARMA model. Monti's test provides numerical evidence that the independence assumption is reasonable, making sensible the methodology and the computations developed in \S \ref{Sec: FDESadd}. In the next session we provide additional numerical evidence, illustrating the good performance of our FDES techniques

\subsection{Testing statistical hypotheses} \label{Sec: SimARFIMA1}

\begin{figure}[h!]
\begin{center}
\begin{tabular}{ccc}
$d^0=0.1$ & $d^0=0.35$ & Power \\ 
\includegraphics[width=0.32\textwidth, height=0.275\textheight, angle=0]{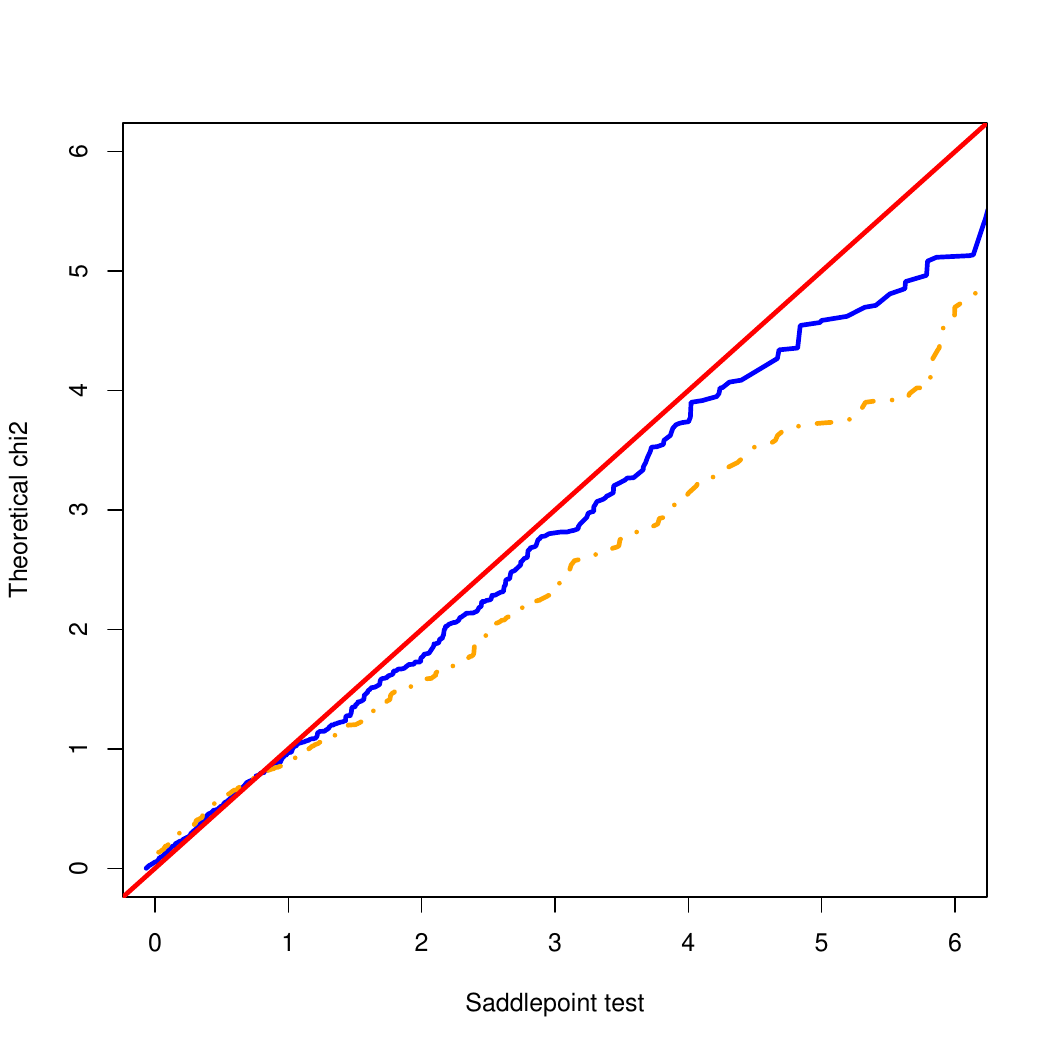} & 
\includegraphics[width=0.32\textwidth, height=0.275\textheight, angle=0]{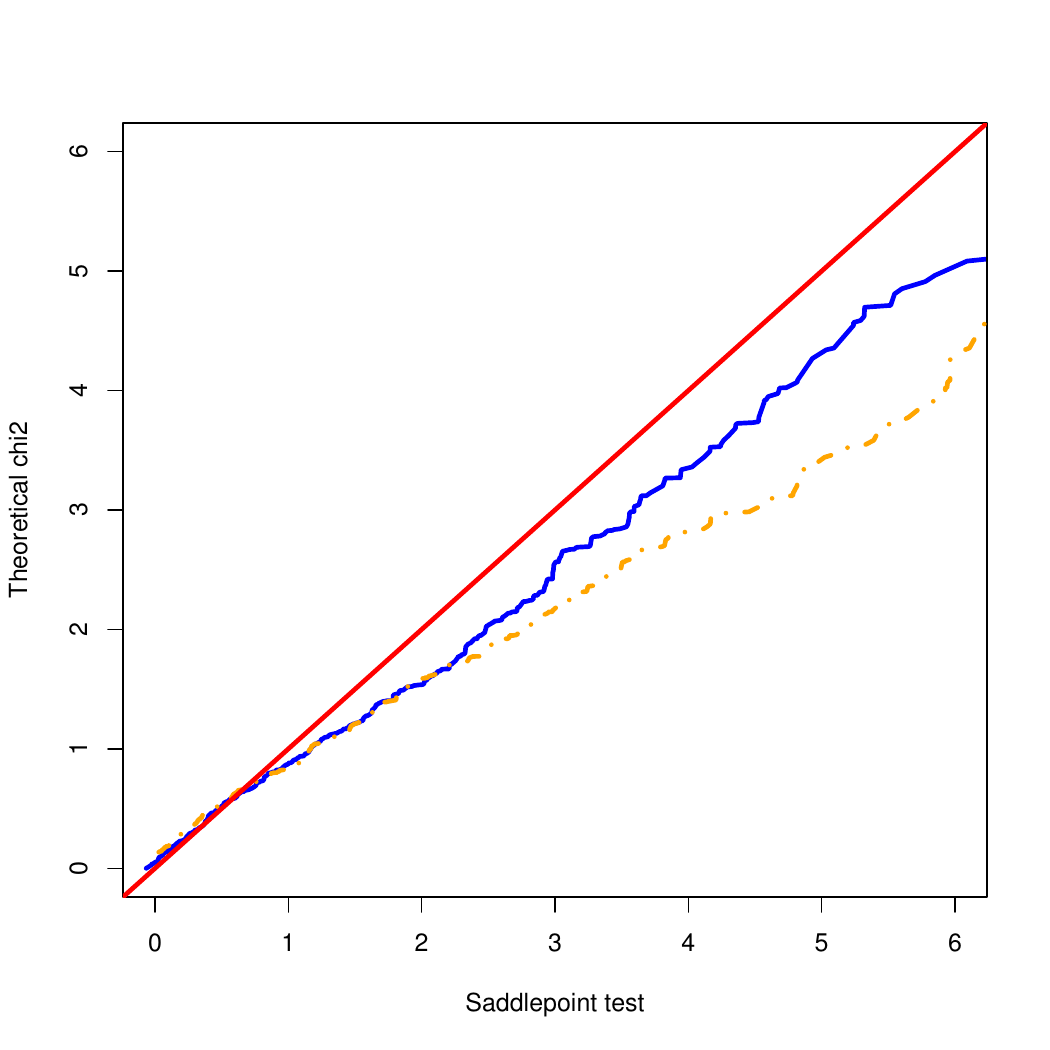} & 
\includegraphics[width=0.32\textwidth, height=0.275\textheight, angle=0]{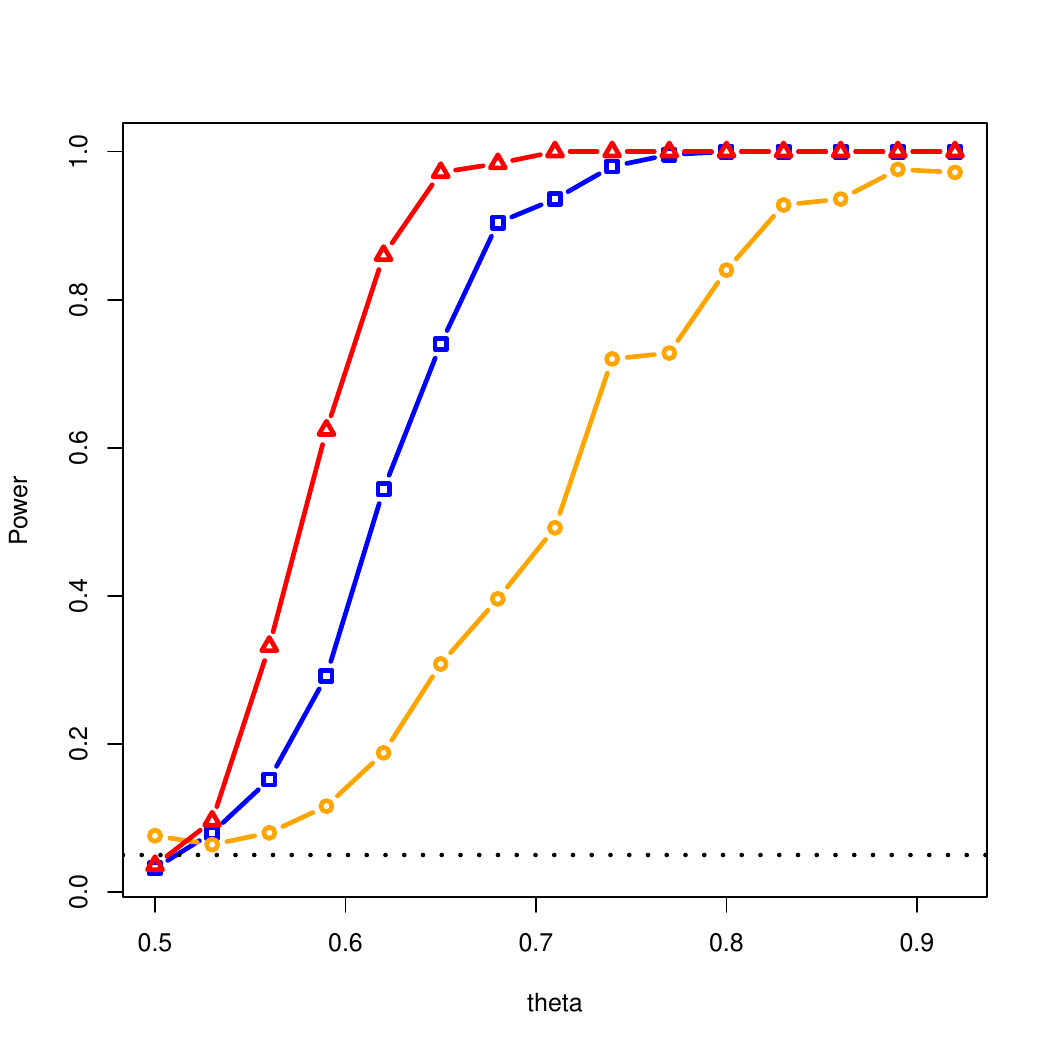} \\
\end{tabular} 
\end{center}
\caption{Gaussian ARFIMA$(0,d,0)$ model. Left and Middle plot: QQ-plot vs the asymptotic  $\chi^2(1)$-distribution of saddlepoint test statistic (\ref{Eq. SADtest1}), for different sample sizes (continuous line $n=250$, dashed line  $n=100$). Right plot: power of the test for $n=100$ (circles), $n=250$ (squares)  and $n=500$ (triangles); dotted horizontal line represents the level at $0.05$.}
    \label{Fig: QQSad1}
\end{figure}

We set the problem of  testing the following null hypothesis about the
long-memory parameter in a Gaussian ARFIMA $(0,d,0)$: 
$\mathcal{H}_0: d=d^0 \quad \text{vs} \quad  \mathcal{H}_1: d>d^0.$ 
We consider this testing problem for different values of $d^0$ and for different sample sizes. Specifically, we set $d^0=0.1$ (moderate long memory) and 
$d^0=0.35$ (strong long memory) and we study the behaviour of the saddlepoint test statistic $\tilde S(\hat \theta_n)$ in (\ref{Eq. SADtest1}), for two small sample sizes: $n=100$ and $n=250$. In Figure \ref{Fig: QQSad1} we display the QQ-plot for the test against his theoretical $\chi_1^2 $ distribution across the different values of $d^0$: the plots illustrate that the saddlepoint test has a distribution which is very close to the theoretical one when $n=250$. The plots for $n=100$ depict the accuracy improvements that the test features, when the sample size moves from $n=100$ to $n=250$, in the presence of both moderate and strong long memory. In line with the outcomes in  Figure \ref{Fig: QQWald}, the Wald test has  a large size distortion  for all the considered sample sizes. For instance, when $n=250$ the 95\%-quantile is 21.78, whilst  the 95\%-quantile of the $\chi_1^2$ is about 3.77. To complete the picture we study the power of the test under local departures (Pitman-type sequence of alternative hypotheses) when the sample size increases. We display the results in the right plot of Figure \ref{Fig: QQSad1}: the power curves illustrate that the test has non trivial power for the local alternatives, displaying a remarkable improvement when $n$ goes from $100$ to $250$.

Now, let us consider the test procedure  in Algorithm \ref{algo1} to perform a test on an ARFIMA($0$,$d$,$0$) for $\mathcal{H}_0 : d = 0$ versus $\mathcal{H}_1: \ d \neq 0$. We compute the true distribution of $\tilde{W}_n(\theta^0)$ for different sample sizes using $10^4$ ARFIMA($0$,$d$,$0$) Monte Carlo replicates of length $n=30$ and $n = 250,$ with $d=0.$ Then, for the first $R = 250$ of these generated time series, we run an importance sampling algorithm similar to Algorithm \ref{algoIS} to obtain the saddlepoint approximation of the full c.d.f.\ of $\tilde{W}_n(\theta^0).$ As a representative c.d.f., we take the functional median of these $R$ c.d.f., and we generate the QQ-plots of Figure \ref{MC0d0}, comparing the true (Monte Carlo) quantiles of $\tilde{W}_n(\theta^0)$ with the ones of the median c.d.f.\ We also add the quantiles of the $\chi_1^2$ approximation, to see if the FDES typically improves on them.

\begin{figure}[h!]
\begin{center}\label{MC0d0}
\begin{tabular}{cc}
$n=30$ & $n=250$\\ 
\hspace{-5mm}\includegraphics[scale = 0.4]{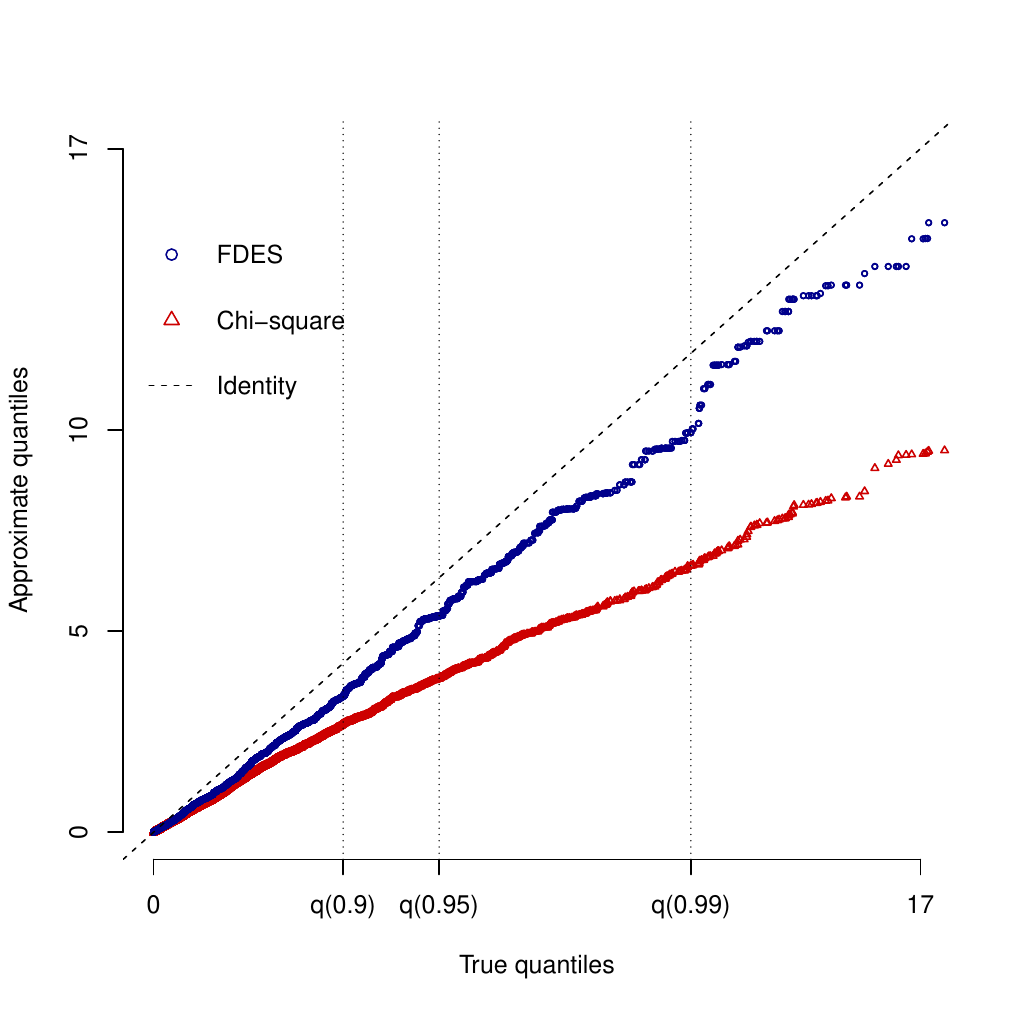}\ &
\includegraphics[scale = 0.4]{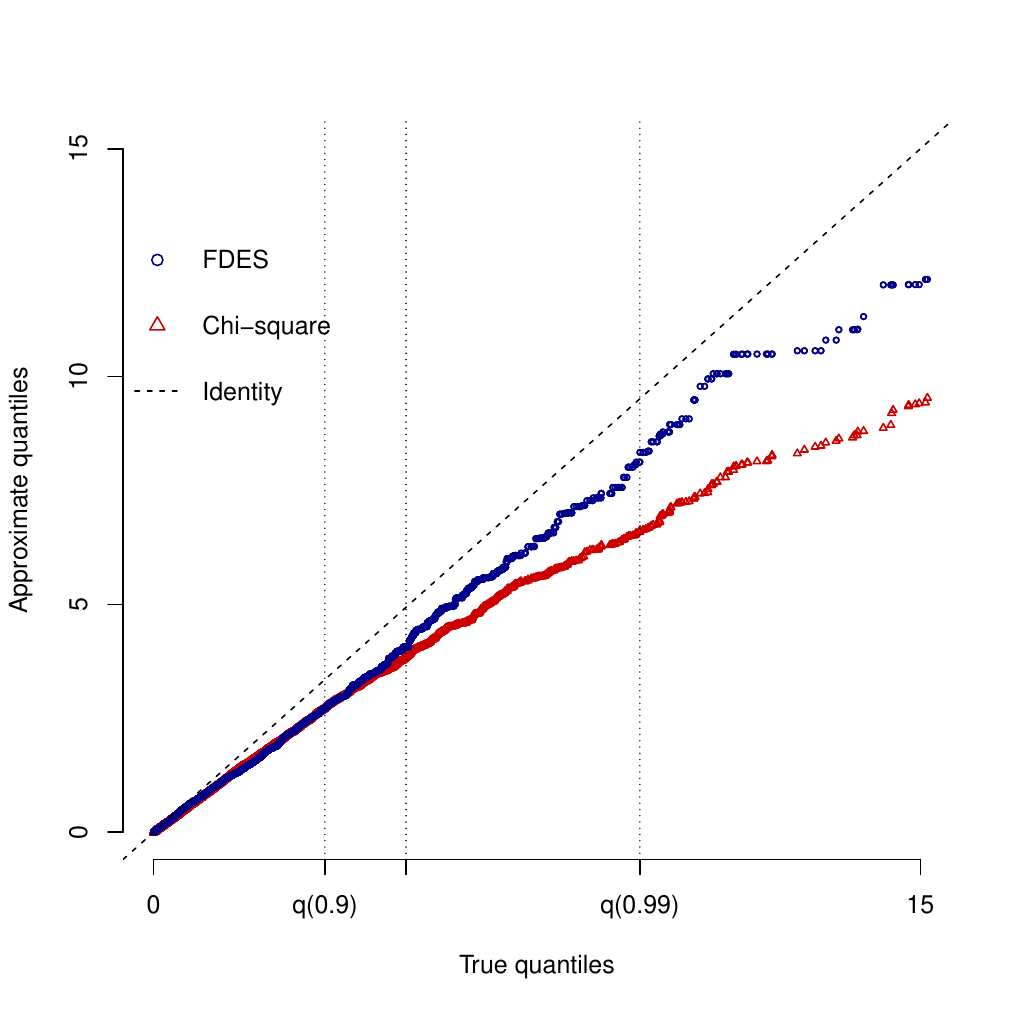}\\
\end{tabular}
\end{center}

\caption{ARFIMA$(0,d,0)$ model $\tilde{W}_n(\theta^0)$ statistic for $\mathcal{H}_0:\ d=0$ versus $\mathcal{H}_1: \ d \neq 0$. For different sample size $n$, the QQ-plots compare the true distribution (as obtained using $10^4$ Monte Carlo runs) of the $\tilde{W}_n(\theta^0)$ statistic to the quantiles of the FDES and asymptotic $\chi_1^2$ distributions. The specific $90 \%$, $95 \%$ and $99 \%$ true percentiles are also highlighted, as the most frequently used in practice.}\label{MC0d0}
\end{figure}

We observe that the characteristic quantiles of the saddlepoint approximation are typically closer to the true quantiles of $\tilde{W}_n(\theta^0)$ than the ones of the $\chi_1^2$ approximation. Both methods converge to the true quantiles as $n$ increases, but the saddlepoint method seems to perform better for all sample sizes. Here we only display the $n=30$ and $n=250$ cases, but we observe the same properties also for other sample sizes.

Let us now consider a more complex process, entailing more dependence in the data; we repeat the same experiment but generating time series from an ARFIMA($1$,$d$,$1$) under $\mathcal{H}_0 : \theta = (0.5, 0.25, 0.5).$ Due to the increase in the dependence and the number of parameters, both methods typically need a larger sample size. We show the Monte Carlo results for $n = 100$ and $n=500$ in Figure \ref{MC1d1} below.

\begin{figure}[h!]
\begin{center}
\begin{tabular}{cc}
$n=100$ & $n=500$\\ 
\hspace{-5mm}\includegraphics[scale = 0.39]{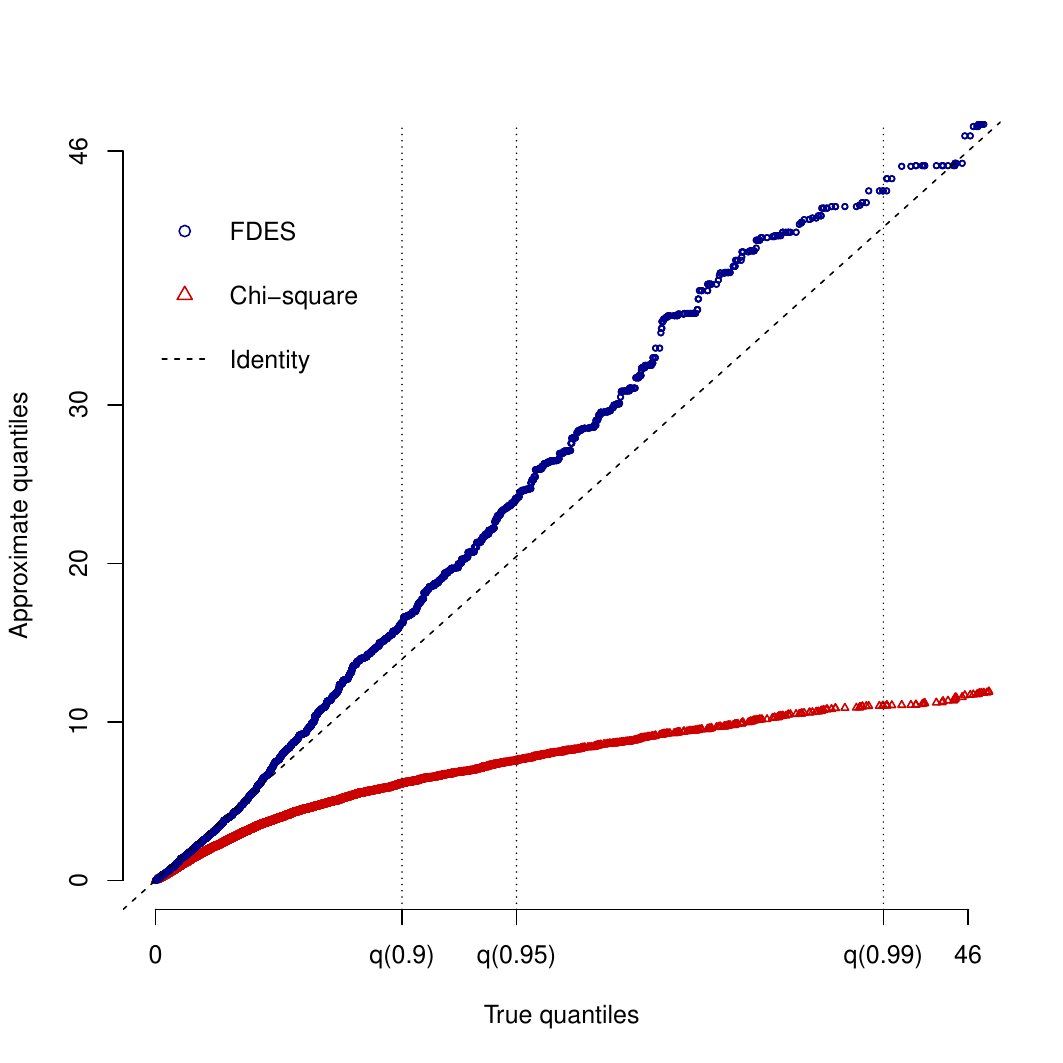}\ &
\includegraphics[scale = 0.39]{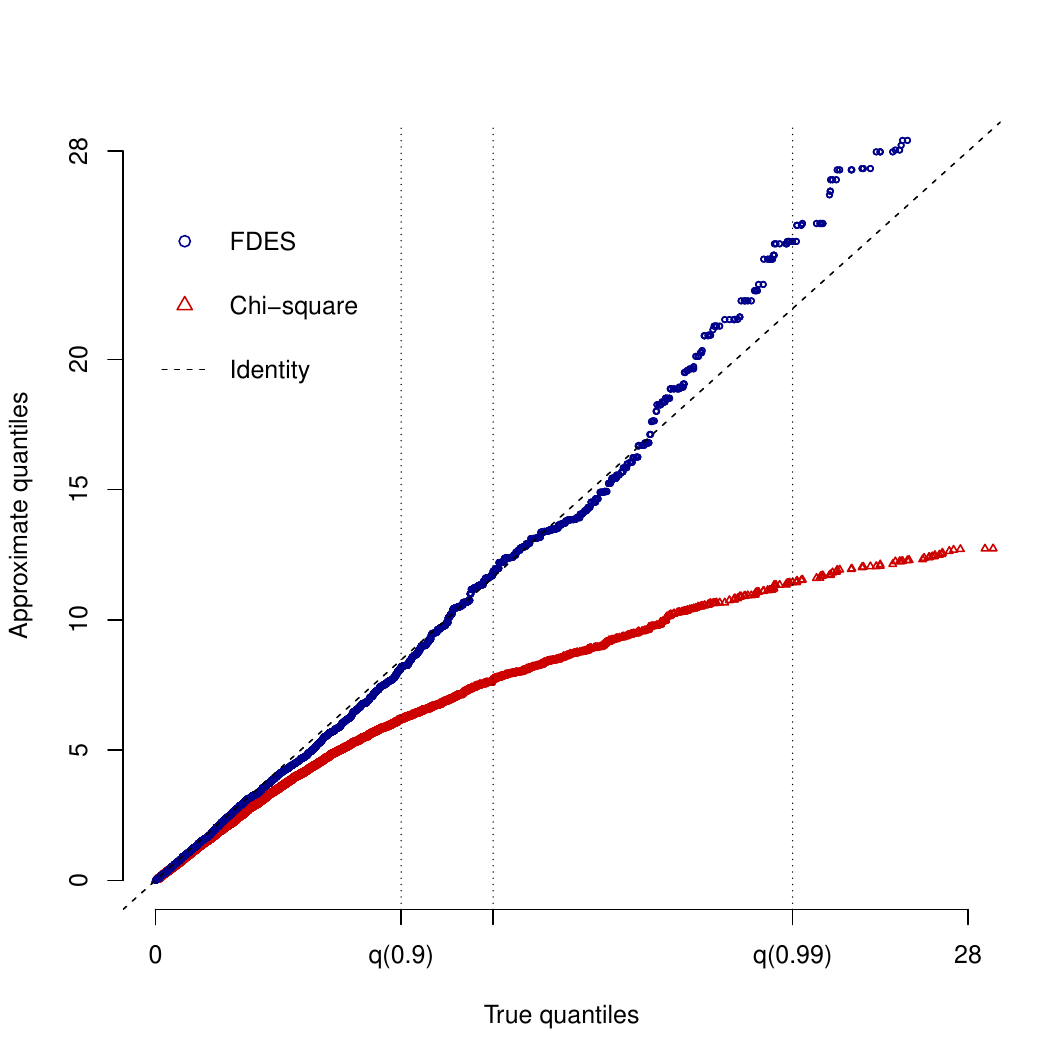}\\
\end{tabular}
\end{center}

\caption{ARFIMA$(1,d,1)$ model $\tilde{W}_n(\theta^0)$ statistic for $\mathcal{H}_0:\ \theta = (0.5, 0.25, 0.5)$ versus $\mathcal{H}_1: \theta \neq (0.5, 0.25, 0.5)$. For different sample size $n$, the QQ-plots compare the true distribution (as obtained using $10^4$ Monte Carlo runs) of the $\tilde{W}_n(\theta^0)$  statistic to the quantiles of the FDES and asymptotic $\chi_3^2$ distributions. The specific $90 \%$, $95 \%$ and $99 \%$ true percentiles are also highlighted, as the most frequently used in practice.}\label{MC1d1}
\end{figure}

Note that the Whittle estimation is numerically less stable for small samples in the ARFIMA$(1,d,1)$ case, with three parameters to estimate. Thus we only report here the Monte Carlo samples which led to convergence, namely about $92.4 \%$ of them for $n=100$  and $99.5 \%$ for $n=500$.

\section{Application to a real dataset}

The Pacific Decadal Oscillation (PDO) index measures a key climatological feature of the Southern hemisphere (\citet{zhang_PDO_1997}). Similarly to El Niño/Southern Oscillation phenomenon, the PDO extremes correspond to episodes of abnormal weather conditions, located in North America and the Pacific Basin. The PDO index is mainly a function of sea level pressures (SLPs) and sea surface temperatures (SSTs). For instance, it takes a positive value if SLPs are below average over the North Pacific, SSTs are high on the Pacific Coast and low in the North Pacific.

\citet{whiting_PDO_2003} show strong evidence of long memory in the PDO index, and model the process with an ARFIMA$(0,d,0)$. Their PDO index data cover the period from 1900 to 1999, giving a Whittle's estimate $d=0.446$: this estimated value  suggests a very strong long memory. In \citet{yuan_PDO_2014}, the authors also model persistence by fractionally integrated processes, adding the more recently available data but discarding the measures made before 1920 for the sake of reliability. To check if these results are still up to date, we make use of our FDES techniques.

We start our data analysis by following \citet{whiting_PDO_2003} and \citet{yuan_PDO_2014}: we aggregate the monthly PDO index\footnote{Available at \url{https://www.ncei.noaa.gov/pub/data/cmb/ersst/v5/index/ersst.v5.pdo.dat}} into an annual time series, from 1920 to 2022, and remove the slight linear trend. We display the resulting data in Figure \ref{Fig: PDO} below, as well as the log-periodogram ordinates against their log-frequencies, hinting at some long memory effect as the slope is negative (see \citet{Beranetal13}). Thus, we estimate an ARFIMA($1$,$d$,$0$) and use the FDES to test (as in Algorithm \ref{algo1}) if the obtained estimates match the ones of previous studies, namely $\mathcal{H}_0: \theta = (\phi \ d) = (0, 0.446).$ We add an autoregressive parameter $\phi$ in our model as it allows to capture the transient effects that the differencing parameter $d$ is missing, by measuring the persistence only.

\begin{figure}[h!]
\begin{center}
\begin{tabular}{c}
\includegraphics[scale=0.6]{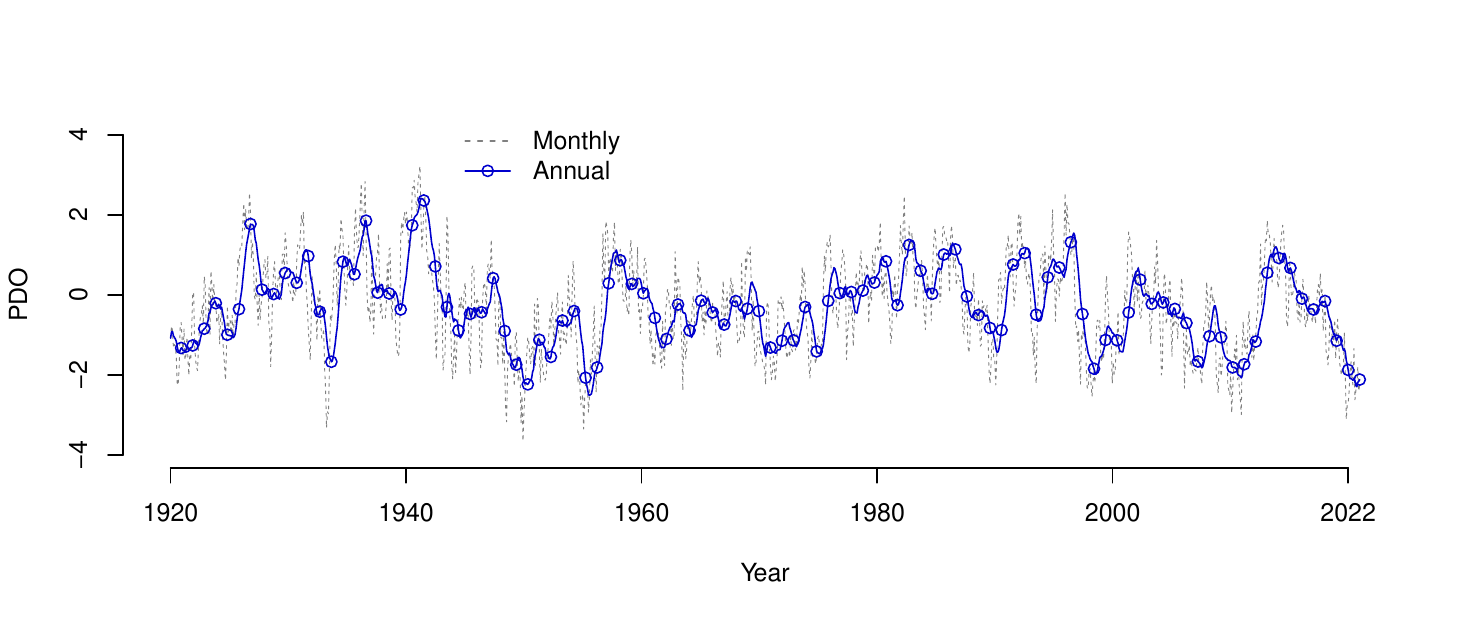}\\
\vspace{-1.6cm}\\
\includegraphics[scale=0.6]{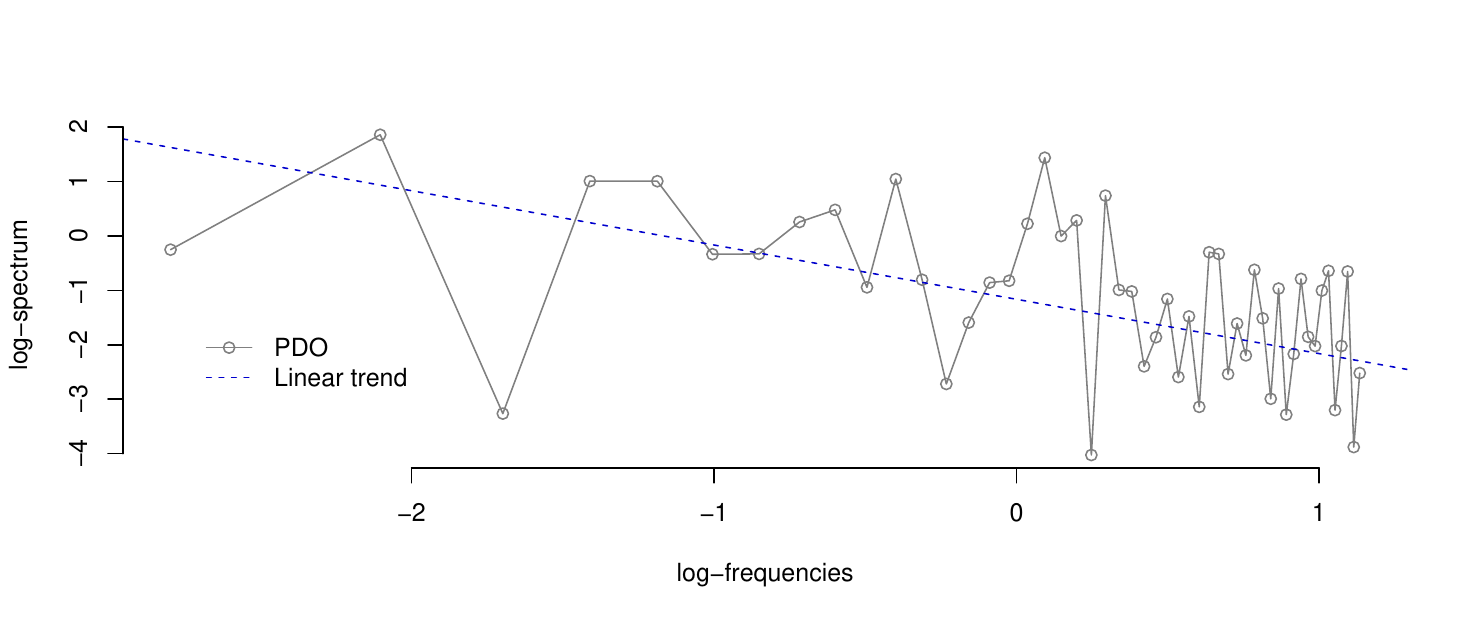}\\
\end{tabular}
\end{center}
\caption{PDO data from 1920 to 2022 (upper plot), and the log-periodogram ordinates against their log-frequencies (lower plot). The slightly negative slope in the last plot hints at some long memory effect.} 
\label{Fig: PDO}
\end{figure}

Following Algorithm \ref{algo1} and treating the variance as a nuisance parameter as in \citet{M97}, we define the Whittle estimating functions for the ARFIMA($1$,$d$,$0$) as in \eqref{Eq. profiled}  
for $j= 1,\ldots,51.$ Solving \eqref{Eq: Mestimator}, the estimate of the parameters is $\widehat{(\phi \ d)} = (0.448,0.088),$ the FDET statistic (as in the left-hand side of \eqref{QuadFHat}) is $5.718$ and the Wald statistic (as in \eqref{wald}) is $4.598.$ In Table \ref{Tab: App} below, we give the results of both the FDES and Wald test procedures.

\begin{table}[h!]
\caption{Quantiles of the test statistics, by the FDES and the $\mathcal{X}^2_2$ approximations.}
\begin{center}\label{Tab: App}
\begin{tabular}{l|ccc}
\backslashbox{\hspace{-0mm}Method.}{\hspace{-8mm}Quant.} & $0.9$ & $0.95$ & $0.99$\\ \hline
\vspace{-3mm}\\
FDES & $3.695$ & $5.471$ & $9.891$\\
$\mathcal{X}^2_2$ & $4.605$ & $5.991$ & $9.210$
\end{tabular}
\label{testPDO}
\end{center}
\end{table}
\noindent In the case of the FDET statistic, we observe that the FDES test allows to reject $\mathcal{H}_0: (\phi, d) = (0,0.446)$ at a $5 \%$ level, as opposed to the $\mathcal{X}^2_2$ approximation. The Wald statistic allows to reject only at a confidence level of $10 \%$ together with the FDES approximation.

\begin{figure}[h!]
\caption{Non-rejection regions of the FDET statistic at the $95 \%$ level.}
\vspace{-8mm}
\begin{center}
\includegraphics[scale=0.5]{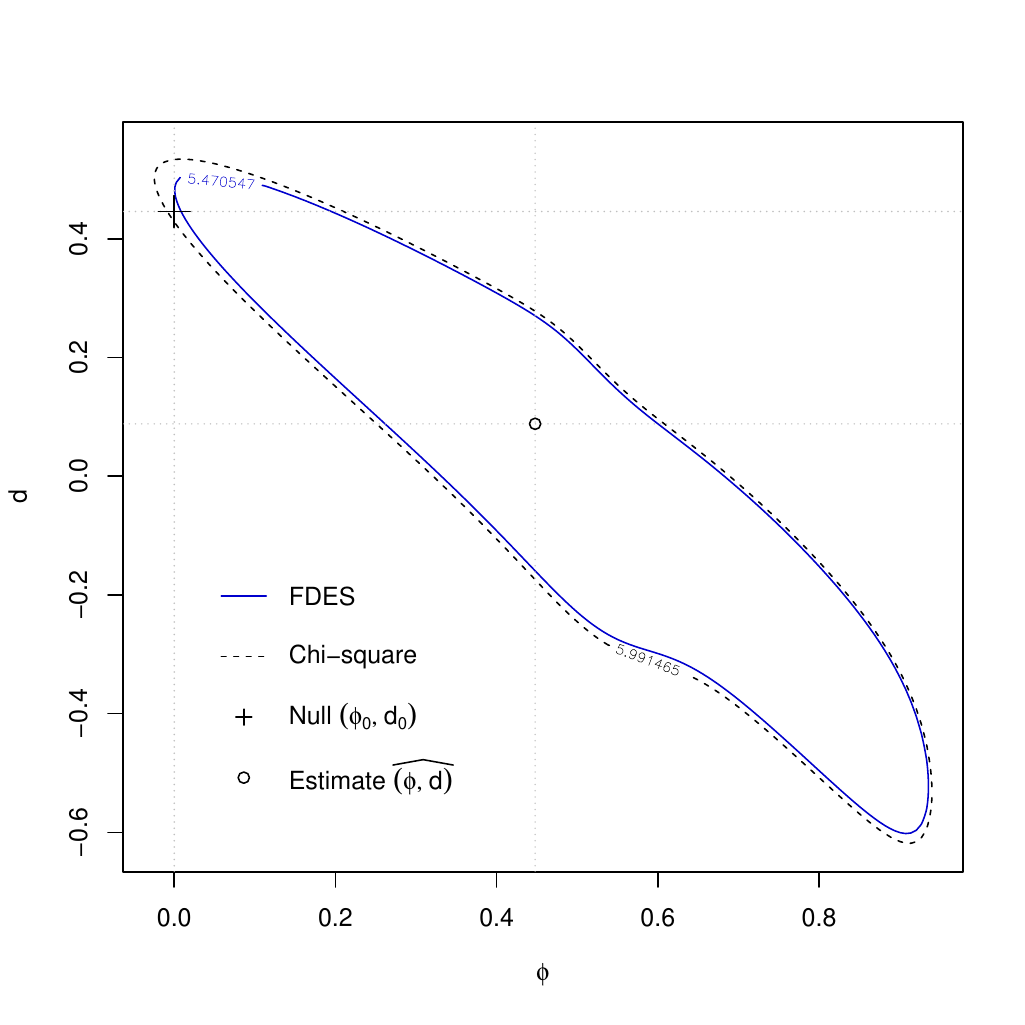}\\
\end{center}
\caption*{We observe that the null hypothesis lies in between the FDES and the $\mathcal{X}^2_{2}$ $95 \%$ level curves.}
\label{PDORegions}
\end{figure}

Thus, the first statistical issue we may flag is the non-rejection of the Wald test with the $\mathcal{X}^2_{2}$ approximation, which is likely to be a wrong conclusion in this setting. The second one, from the rejection of $\mathcal{H}_0: (\phi, d) = (0,0.446)$ by the FDES procedure, is the apparent non-stationarity of the PDO time series, which requires some further investigations. As a summary of our results, Figure \ref{PDORegions} below represents the non-rejection regions of the FDET statistic at the $95 \%$ level, as estimated by the FDES and the $\mathcal{X}^2_{2}$ approximation.

\section{Conclusion}

We introduce a novel class of saddlepoint techniques for time series data analysis. Our results complete the toolkit already defined in \cite{LVR19} and have several connections with other extant frequency domain techniques, routinely applied in econometrics and statistics (namely, the FDB, EL, and ET). These connections are methodologically relevant not only because they open the door to the definition of an holistic approach to the frequency domain time series data analysis, but also because they allow us to define a novel and easy-to-implement FDES procedure for testing statistical hypotheses, in SRD and LRD time series. We hope that our unified view and treatment of saddlepoint approximations will trigger the interest of statisticians and econometrician, yielding additional findings in the frequency domain analysis of time series data.

Among the possible novel research directions that our methodology can stimulate, we conjecture that additional results on the link between FDEL and FDES can be obtained by bounding the $\alpha$-mixing coefficients of the periodogram ordinates, as in \cite{MMmixing}. In particular, this may help to control the error term in \eqref{Eq. MYFDEL}. Moreover, it may open the door to the use of the whole GEL family described in \cite{kakizawa13} which can be the building block for density approximations in the spirit of \cite{BN1983}. Essentially,  the idea is to define a generalization of the extant Esscher titling, which is based on the minimization of the Kullback-Leibler (KL) divergence between the actual density and the tilted density (see e.g. \cite{LVRI23}), and minimize any other divergence belonging to the Cressie-Read family. In the same vein,  we plan to consider distances based on  optimal transport  theory (see \cite{V08}), e.g.\ the $p$-Wasserstein distance as in \cite{Blanchet_Kang_Murthy_2019} and explore their advantages (robustness) with respect to the use of KL. Finally, we believe that the importance sampling approach described in Algorithm \ref{algo1} and Algorithm \ref{algoIS} may be suitably modified to deal with
 testing problems for other stochastic processes (beyond the time series case discussed in this paper), like testing hypotheses in the presence of nuisance parameters in the class of spatial autoregressive processes for panel data considered in \cite{JLRS23}.


\bibliographystyle{apalike}
\bibliography{biblio}
\newpage
\appendix
\input{Appendix}

\end{document}

%% file: Appendix.tex
\section{Exponential-based techniques} \label{App: FEXP}

\subsection{FEXP processes and GLM} \label{Sec_FEXP}

\cite{B93} and \cite{PR95} introduced the fractional exponential (FEXP) processes a large class of stochastic processes, which are able to model both short- and 
long-memory. FEXP processes represent  an extension of \citet{Bloom73}. The advantage of this class is that one can estimate the parameters by using widely applied methods for generalized linear models (GLM), which are readily available in the many statistical software packages.

To set up the notation Let us define $q_0\equiv 1$ and let $q_1,q_2,...,q_p$ be symmetric functions which are piecewise continuous in the interval $[-\pi, 
\pi]$. For any $n \in {N}$ we set
$m=\lfloor (n-1)/2 \rfloor$ and let be $\theta$ the $(p+2)$-dimensional vector
$
\theta=(-2d,\eta_0,\eta_1,...,\eta_p)^T,
$ 
with  $0\leq d <0.5$. Then, $\{X_t\}$ is called a FEXP process with short-memory components $q_1,...,q_p$ and long-memory component $g$, if its spectral density is given
by:
\begin{equation}
f(\lam;\theta)=\vert 1-g(\lam) \vert ^{-2d}\exp\left\{  \sum_{k=0}^{p} \eta_k q_k(\lam) \right\},
\label{Eq: FEXPspd}
\end{equation} 
for $g:\Pi \rightarrow \mathbb{R}^{+}$  even function such that
$\lim_{\lambda\rightarrow 0} g(\lambda)/\lambda=1$.
The spectral density in (\ref{Eq: FEXPspd}) has the form as in (\ref{Eq: gen_spec}) and it can be parameterized in terms of the Hurst exponent $H=d+1/2$:
when $H=0.5$ ($d=0$) we have SDR, whereas for $H\in (0.5,1)$ ($d\in(0,0.5)$), we have LRD. In its generality, this class is able to approximate 
with arbitrary accuracy any piecewise continuous spectral density. 

For FEXP processes, \cite{B93} defines a frequency domain
$M-$estimator and estimates the parameter $\theta$ 
in a GLM context, with underlying exponential distribution and noncanonical link function. 
An estimator of $\theta$ is 
defined by (\ref{Eq: Mestimator}) with the estimating 
function as in (\ref{Eq: psiM}) and covariates (not depending on $\theta$)
\begin{equation}
\label{Eq: zberan}
z^T_j=[\ln g
(\lambda_{j}), 1,...,q_p(\lambda_{j})]^T.
\end{equation}

Making use of (\ref{Eq. exp1}), we write the following multiplicative model for the periodogram ordinates:
$I_{j} =  f(\lambda_{j};\theta) \zeta,$ for every $j$,
where the $\zeta_j's$ are independently and identically distributed exponential random variables, with mean one; 
see, e.g., \cite{B93} and reference therein. 

In the FEXP setting, Whittle's estimator can be obtained by GLM M-estimation and we may derive small sample techniques making use 
of the saddlepoint methods available for multivariate M-estimators in the (i.)i.d. setting. 
However, the available results need to be adapted to our frequency domain setting. Specifically, for $f(\lam_{j};\theta)$ as in (\ref{Eq: FEXPspd}) and for $\theta^0$ specified by $\mathcal{H}_0$ in (\ref{Eq. TestProblem}),  we have
that, 
for the M-estimator  defined by (\ref{Eq: Mestimator}) and (\ref{Eq: psiM}),  
the cumulant generating function of $\psi_j(I_j;\theta)$ in (\ref{Eq: Kpsij}) is
\begin{eqnarray}
\label{Eq: cumul_gen_fct}
K_{\psi_j}(\upsilon;\theta)=\ln \left(-\frac{1/f(\lam_{j};\theta^0) }{\frac{\upsilon^T z_j}{f(\lam_{j};\theta)}-\frac{1}{f(\lam_{j};\theta^0)} }\right)-\upsilon^T z_j.
\end{eqnarray}

\subsection{ARFIMA(0,$d$,0)}

Let us consider the Gaussian ARFIMA(0,$d$,0),  
having dynamics as in (\ref{Eq: FARIMAG}). The spectral density at $\lam$ is as in (\ref{Eq. FARIMAsp}). Suppose we want to test the null hypothesis of the Monte Carlo exercise in \S \ref{Sec: SimARFIMA1}: $\mathcal{H}_0: d=d^0$ vs  $\mathcal{H}_1: d>d^0$. To proceed further, we notice that the FARIMA$(0,d,0)$ belongs to the
FEXP family; see \citet{B93}. Indeed, looking at (\ref{Eq: FEXPspd}) and setting $g(x)=\vert 1-\exp(i\lam)  \vert$ and $\eta_k\equiv 0$ for $k>0$, we obtain
the ARFIMA spectral density in (\ref{Eq. FARIMAsp})---multiplied by $2\pi$. Thus a GLM estimator is defined using $z_j=\ln g(\lambda_j)$ in (\ref{Eq: psiM}) and $K_{\psi_j}(\upsilon;d)$ is obtained as in (\ref{Eq: cumul_gen_fct}), with $d=d^0$.

\section{Empirical saddlepoint and empirical likelihood in the frequency domain: unexplored connections}  \label{AppC}

In this Appendix, for the ease of notation, we write $\psi^{(k)}_j (\theta)$ for $\psi^{(k)}_j (I_j,\theta)$. We explain how to combine the results available
in \cite{RM93} and in \citet{M97}, to connect FDES  and FDEL.

As in \cite{RM93}, we consider  the $(\text{dim}(\Theta) \times \text{dim}(\Theta))$-matrix $B(u)$ ($\text{dim}(\Theta)$ represents the dimension of the parameter
space), whose elements are
\begin{equation}
\{B(u)\}_{k,l} = \frac{1}{m} \sum_{j=1}^{m} \sum_{r=1}^{\text{dim}(\Theta)} \left\{  \psi^{(k)}_j (\hat\theta_n) \frac{\partial \psi^{(l)}_j}
{\partial \theta_r} \Big\vert_{\hat\theta_n} + \psi^{(l)}_j (\hat\theta_n) \frac{\partial \psi^{(k)}_j}{\partial \theta_l}\Big\vert_{\hat\theta_n}  \right\} 
u_r,
\end{equation} 
and define also the vector $c(u)\in {R}^{\text{dim}(\Theta)}$:
\begin{equation}
c(u)=\frac{1}{m} \sum_{j=1}^{m} \sum_{k=1}^{\text{dim}(\Theta)} \sum_{l=1}^{\text{dim}(\Theta)} \left\{ \frac{\partial^2 \psi_j(\theta)}
{\partial \theta_k \partial \theta_l}\Big \vert_{\hat\theta_n} u_k u_l \right\}.
\end{equation}
\noindent As in \citet{M97}, we let the FDEL solve the system of estimating equations (\ref{Eq: EL})
and in Monti's paper (see (4.8) on page 398) an expansion of $\hat\xi$  up to the term of order $O_P(m^{-1})$ is available. 
For our purposes, we need to  compute the same kind of expansion up to the order $O_P(m^{-3/2})$ and evaluate it at $\hat\theta_n$. To this end, following the calculation in \cite{RM93}, we get, for $\theta=\hat\theta_n + m^{-1/2} u$, 
\begin{eqnarray}
\hat\xi &=& m^{-1/2} \hat\Sigma^{-1} \hat M u + \frac{m^{-1}}{2} \hat\Sigma^{-1}_{\psi_j} c(u)  - m^{-1} \hat\Sigma^{-1} B(u) \hat\Sigma^{-1} \hat M u  \notag \\ 
&+& m^{-2} \sum_{j=1}^{m} \left\{ u^T \hat M \hat\Sigma^{-1} \psi_j(\hat\theta_n)  \right\}^2 \hat\Sigma^{-1} \psi_j(\hat\theta_n) + O_P(m^{-3/2}) 
\label{Eq: ExpEL},
\end{eqnarray}
where the first term implies also that $\hat\xi=O_P(m^{-1/2})$. We consider Owen's statistic
$\hat{W}(\theta)$ 
 and we  replace (\ref{Eq: ExpEL}) in the Taylor expansion of $\hat{W}(\theta)$, obtaining:
\begin{eqnarray}
\hat{W}(\theta) &=& u^T \hat M^{-1} \hat\Sigma^{-1} \hat M^{-T} u - m^{-1/2} u^T \hat M^{T} \hat\Sigma^{-1} B(u) \hat\Sigma^{-1} \hat M u \notag \\ & + & m^{-1/2} u^T \hat M^{T} \hat\Sigma^{-1} c(u) + \frac{2m^{-3/2}}{3}  \sum_{j=1}^{m} \left\{ u^T \hat M^T \hat\Sigma^{-1} \psi_j(\hat\theta_n)   \right\}^3 \nonumber \\
& + & O_P(m^{-1}). \label{Eq: OwenExp}
\end{eqnarray}

\noindent Now, let us turn to the empirical saddlepoint $\hat{\upsilon}(\theta)$, under the exact independence of the periodogram ordinates. 
To bridge the FDES density approximation  of \citet{LVR19} and Monti's FDEL, we drop the assumption that the distribution of the periodogram ordinates is an exponential, but we keep their independence. To this end, we combine the construction in \citet{RM93, M97} with the one of  \citet{RW94}). We notice that $\psi_j\left(\theta\right)$ contains the terms $(I_i/f({\lambda_i;\theta})-1)$ interpretable 
as \textit{Pearson-type residuals}---namely, pivotal quantities with zero mean and unit variance. This interpretation is 
common in the frequency domain literature (we refer to \citet{FH92}, \citet{B93} and \citet{DJ96}, among the others) and it is based on 
a re-centering and re-scaling of the periodogram ordinates, as obtained exploiting the knowledge of the moments of their asymptotic distribution.
Then the equation defining the empirical saddlepoint is
\begin{equation}
\sum_{j=1}^{m} \psi_j(\theta) \exp\{ {\hat\upsilon^T} \psi_j(\theta)\} =0,
\end{equation}
therefore, from \citet{RW94} we obtain the empirical saddlepoint approximation in Equations \eqref{Eq: empiricalSAD}---\eqref{Eq: aempiric1}.
Then we have, for any $u$ in compact set,
$f_{\hat\theta_n}(\theta^0+m^{-1/2}u)=\hat{g}_{\hat\theta_n}(\hat\theta_n + m^{-1/2}u)\{1+m^{-1/2} a_m(u)+O_p(m^{-1})\},
$ 
as $m \rightarrow \infty$, where $\hat a_m(u)=O_p(1)$ is defined in \citet{RW94}.
From (\ref{Eq. Lemma}) it follows that $\upsilon(\theta)=O(n^{-1/2})$, for $\theta$ in a root-$n$ neighbourhood  of $\theta^0$. Thus, using a first order Taylor expansion (in $\hat\upsilon$)  of the saddlepoint equation we obtain
\begin{equation}
0 = m^{-1}\sum_{j=1}^{m} \psi_j(\theta) \exp\{\hat\upsilon^T \psi_j(\theta)\}= m^{-1}\sum_{j=1}^{m} \psi_j(\theta) \left\{1 +\hat\upsilon^T \psi_j(\theta)+ O (m^{-1}) \right\},
\end{equation}
 which yields
\begin{equation}
m^{-1}\sum_{j=1}^{m} \psi_j(\theta) \left\{1 +\hat\upsilon^T \psi_j(\theta)\right\}=O_p(m^{-1}).
\label{Eq: Taylor1st}
\end{equation}
From this point on, the calculation is similar to the development available in \cite{RM93}. Thus, using (\ref{Eq: Taylor1st}), we write $\hat\upsilon(\theta) = \upsilon_1(\theta) +\upsilon_2(\theta)$, where we label $\upsilon_1
(\theta)=m^{-1/2} \hat\Sigma^{-1} \hat M u$ and $\upsilon_2(\theta)=O(m^{-1})$. A second order Taylor expansion 
of the saddlepoint equation, yields
$$
m^{-1}\sum_{j=1}^{m} \psi_j(\theta) \left[1 +\{\hat\upsilon_1^T + \hat\upsilon_2^T\} \psi_j(\theta) + \frac{1}{2} \{\hat\upsilon_1^T \psi_j(\theta) \}^2 \right]=O_P(m^{-3/2}),
$$
which implies 
\begin{eqnarray}
\hat\upsilon_1 (\theta) + \hat\upsilon_2 (\theta) &=& -m^{-1/2} \hat\Sigma^{-1}  \hat M u - \frac{m^{-1}}{2}  \left[ m^{-1} \sum_{j=1}^{m} \{u^T \hat M^{T} \hat
\Sigma^{-1} \psi_{j}(\hat\theta_n) \}^2  \hat\Sigma^{-1} \psi_j(\hat\theta_n)  \right] \notag \\
&+& m^{-1} \hat\Sigma^{-1} B(u) \hat\Sigma^{-1} \hat M u -\frac{m^{-1}}{2} \hat\Sigma^{-1} c(u) + O_P(m^{-3/2}). \label{Eq: SaddExpansion}
\end{eqnarray}
A  Taylor expansion of the cumulant generating function $\hat K(\theta)$ yields
\begin{eqnarray}
\ln\left(m^{-1} \sum_{j=1}^{m} \left[ 1 + (\hat\upsilon_1^T + \hat\upsilon_2^T) \psi_j(\theta) + \frac{1}{2} \{(\hat
\upsilon_1^T + \hat\upsilon_2^T) \psi_j(\theta) \}^2 + \frac{1}{6} \{(\hat\upsilon_1^T + \hat\upsilon_2^T) \psi_j(\theta) \}^3   
\right] + O_P(m^{-2}) \right) \nonumber \\
= m^{-1} \sum_{j=1}^{m} \left[ 1+  \hat\upsilon^T \psi_j(\theta_n,\theta) + \frac{1}{2} \{\hat
\upsilon^T \psi_j(\theta_n,\theta) \}^2 + \frac{1}{6} \{\hat\upsilon^T  \psi_j(\theta_n,\theta) \}^3   
\right] + O_P(m^{-2}), 
\label{Eq. 2TaylorKcheck}
\end{eqnarray}
To proceed further, we
 Taylor expand the term $m^{-1}\sum_
{j=1}^{m} \psi_j(\theta)$ around $\hat\theta_n$ and, making use of (\ref{Eq: SaddExpansion}),  we obtain
\begin{eqnarray}
\hat K(\theta)&=& -\frac{m^{-1}}{2} u^T \hat\Sigma^{-1} \hat M \hat M^T  \hat\Sigma^{-1} u +  \frac{m^{-3/2}}{2} u^T \hat M  \hat\Sigma^{-1} B(u) 
 \hat\Sigma^{-1}   \hat M u \notag \\
 &-& \frac{m^{-3/2}}{2} u^T \hat M \hat\Sigma^{-1}  c(u) - \frac{m^{-5/2}}{6}\sum_{j=1}^{m} \left\{ u^T \hat M^T \hat\Sigma^{-1} \psi_j(\hat\theta_n)   
\right\}^3 + O_P(m^{-2}). \label{Eq: KExp}
\end{eqnarray}

Now, comparing (\ref{Eq: OwenExp}) and (\ref{Eq: KExp}), we get
\begin{equation*}
m \hat K(\theta) = -\frac{1}{2} \hat W(\theta) + \frac{m^{-3/2}}{6} \sum_{j=1}^{m} \left\{ u^T \hat M^T \hat\Sigma^{-1} \psi_j(\hat\theta_n)   
\right
\}^3 + O_P(m^{-1}).  
\end{equation*}
Since $n$ is proportional to $m$, the result in \eqref{Eq. MYFDEL} follows, under the  independence of the periodogram ordinates.